\documentclass{article}
\usepackage[preprint]{neurips_2023}
\pdfoutput=1
\PassOptionsToPackage{numbers, compress}{natbib}

\usepackage[utf8]{inputenc} 
\usepackage[T1]{fontenc}    
\usepackage{hyperref}       
\usepackage{url}            
\usepackage{booktabs}       
\usepackage{amsfonts}       
\usepackage{nicefrac}       
\usepackage{microtype}      
\usepackage{xcolor}         
\hypersetup{
    colorlinks,
    linkcolor={red!50!black},
    citecolor={blue!50!black},
    urlcolor={blue!80!black}
}
\usepackage{nccmath}
\usepackage{xspace}
\usepackage{multirow}
\usepackage{nicefrac}
\usepackage{subfig}
\usepackage{soul}
\usepackage{float}
\usepackage{./formatting/macros}
\usepackage{./formatting/results}
\usepackage{balance}
\usepackage{graphicx}
\usepackage{wrapfig}
\usepackage{arydshln}

\title{Ad-Rec: Advanced Feature Interactions to Address Covariate-Shifts in Recommendation Networks}

\author{\vspace{-0.1in}Muhammad Adnan$^\star$ Yassaman Ebrahimzadeh Maboud$^\star$ Divya Mahajan$^\dagger$ Prashant J. Nair$^\star$\And\texttt{\{adnan,yassaman,prashantnair\}@ece.ubc.ca}  \quad\quad\quad\texttt{divya.mahajan@microsoft.com}\And The University of British Columbia$^\star$\quad\quad\quad\quad\quad\quad\quad\quad\quad\quad\quad\quad Microsoft$^\dagger$
}

\begin{document}

\maketitle

\begin{abstract}
Recommendation models are vital in delivering personalized user experiences by leveraging the correlation between multiple input features. However, deep learning-based recommendation models often face challenges due to evolving user behaviour and item features, leading to covariate shifts. Effective cross-feature learning is crucial to handle data distribution drift and adapting to changing user behaviour. Traditional feature interaction techniques have limitations in achieving optimal performance in this context.

This work introduces \trec{}, an advanced network that leverages feature interaction techniques to address covariate shifts. This helps eliminate irrelevant interactions in recommendation tasks. \trec{} leverages masked transformers to enable the learning of higher-order cross-features while mitigating the impact of data distribution drift. Our approach improves model quality, accelerates convergence, and reduces training time, as measured by the Area Under Curve (AUC) metric. We demonstrate the scalability of \trec{} and its ability to achieve superior model quality through comprehensive ablation studies.

\end{abstract}
\section{Introduction}
\label{sec:introduction}

Recommendation models are essential for delivering personalized recommendations in various web services \citep{matrixfacreco}. Over time, these models have evolved from conventional collaborative filtering designs \citep{matrixfacreco} to deep learning-based approaches \citep{neuralcf, dlrm, MTWnD, tbsm}, leveraging their capacity to capture complex patterns and improve recommendation quality. However, deep learning-based recommendation models face challenges due to covariate shifts caused by dynamic user behaviour and evolving item features \citep{dlrm, tbsm, DeepFM}. These shifts lead to a misalignment between the training and testing data distributions, resulting in degraded performance and limited generalization. This paper aims to address these challenges.

\begin{figure*}[t]
\centering
\subfloat[Non-Sequential Recommendation Model]{
\begin{minipage}[t]{0.4\textwidth}
\includegraphics[width=\textwidth]{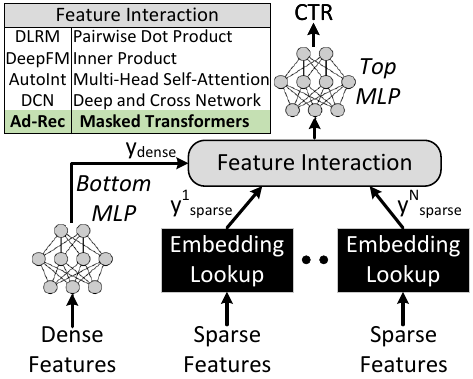}
\vskip -0.1in
\label{fig:recmodel}
\end{minipage}}\hfill
\subfloat[Sequential Recommendation Model]{
\begin{minipage}[t]{0.45\textwidth}
\includegraphics[width=\textwidth]{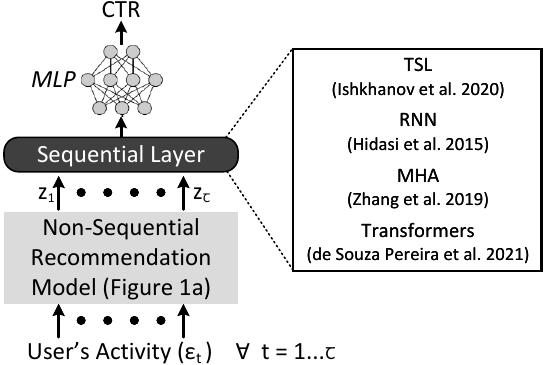}
\vskip -0.1in
\label{fig:seqrecomodel}
\end{minipage}}
\caption{(a) Non-Sequential Deep Learning based Recommendation Model with multiple feature interaction techniques. (b) Sequential recommendation model with multiple sequential layers.}
\vskip -0.15in
\end{figure*}

Deep learning-based recommendation models, shown in Figure~\ref{fig:recmodel}, comprise neural networks, embedding lookup, and, most importantly, feature interactions. While neural networks capture continuous user-related inputs such as timestamps and age to model temporal dynamics, the critical aspect lies in feature interactions. Feature interactions integrate latent representations from both continuous and categorical inputs, allowing the models to generate personalized recommendations. Non-sequential recommendation models focus solely on user activity to generate personalized recommendations, while sequential and session-based models, shown in Figure~\ref{fig:seqrecomodel}, leverage users' historical interactions. Sequential and session-based models gain valuable insights into user preferences by considering the order and context of past actions, thereby enhancing recommendation accuracy.

The feature interaction layer plays a crucial role in recommendation tasks by augmenting non-sequential and sequential models with valuable information beyond individual sparse and dense features \citep{dlrm, tbsm, DeepFM}. However, covariate shifts pose significant challenges for deep learning-based recommendation models. Manually identifying cross-features becomes impractical when dealing with a large number of features. This hinders model generalization. To overcome these challenges, deep neural network (DNN) based feature interaction techniques have emerged \citep{WnD, DeepFM, xDeepFM, Deep_Crossing, autoint, BST, fignn}. These techniques allow for the extraction of higher-order features and effective generalization. However, including irrelevant feature interactions can introduce noise and lead to overfitting \citep{AFM, AIM, khawar2020autofeature, liu2020autofis, su2021detecting}. Moreover, modelling all feature interactions in the same space limits generality. Such an approach also fails to capture diverse patterns.

This paper proposes \trec{}, a masked transformer-based approach to address covariate shifts, eliminate irrelevant cross-features, and encompass various feature interaction patterns. \trec{} incorporates three key elements. First, it uses LayerNorm for mitigating internal covariate shifts. Second, it incorporates Multi-head Attention for modelling feature interactions in multiple subspaces, thereby enhancing generalization. Lastly, it employs Attention Masks to eliminate irrelevant cross-features in different subspaces. By stacking multiple transformer encoders, \trec{} effectively captures feature interactions at several (higher and lower) orders, enabling successful recommendations in the presence of covariate shifts.

We conduct experiments across various non-sequential and sequential models to evaluate \trec. On average, \trec achieves the desired AUC target in \dcntrainiterimpr training iterations compared to state-of-the-art DCN-v2~\citep{dcnv2} model. This translates to training speedup of \dcntrainspeedup across seven different models and four real-world publically available datasets~\citep{criteokaggle, criteoterabyte, avazu, alibaba}.
\section{Related Work}
\label{sec:related_work}
\paragraph{1. Enhanced Models:} Recommendation models are essential for delivering personalized user experiences by suggesting items based on the correlation between multiple input features. Approaches such as collaborative filtering (CF)\citep{neuralcf} and matrix factorization (MF)\citep{matrixfacreco, matrix_fact_ranking, tensor_factor} decompose user-item interactions into latent features. However, they have limitations in capturing complex feature interactions due to their reliance on linear combinations of latent features. Item-based neighbourhood methods offer an alternative for recommendation tasks~\citep{neighbourhood, neighbourhood2}.

Deep learning has revolutionized recommendation models, enabling more accurate capturing of intricate feature interactions. Models like Multi-Layer Perceptron (MLP)\citep{WnD, MTWnD} and neural networks\citep{dlrm} have shown promise in capturing and modelling complex feature relationships. Additionally, techniques such as autoencoders \citep{autorec} and Gated Recurrent Units (GRU)\citep{gru4rec} have been employed to account for temporal dependencies in user interactions. To address the temporal nature of user behaviour, sequential recommendation models\citep{tbsm, BST, bert4rec} have been developed, while session-based recommendation models focus on utilizing the user's current session history~\citep{transformers4rec}. These existing methods still face challenges in effectively capturing and modelling the intricate relationships between features.
\paragraph{2. Learning Feature Interactions:} 
Recommendation models have been extensively studied in the machine-learning community. Traditional approaches, such as collaborative filtering~\citep{neuralcf} and matrix factorization~\citep{matrixfacreco, matrix_fact_ranking, tensor_factor}, have been widely used to decompose user-item interactions into latent features. Item-based neighbourhood methods~\citep{neighbourhood, neighbourhood2} have also been employed for recommendation tasks. With the advent of deep learning, various models, including Multi-Layer Perceptron (MLP)~\citep{WnD, MTWnD}, neural networks~\citep{dlrm}, autoencoders~\citep{autorec}, and Gated Recurrent Units (GRU)~\citep{gru4rec}, have been proposed to capture more complex feature interactions and temporal dependencies in user interactions. Sequential recommendation models~\citep{tbsm, BST, bert4rec} and session-based recommendation models~\citep{transformers4rec} have also gained attention in capturing the temporal sequence of user behaviour. While these existing methods have made significant contributions to feature interaction modelling in recommendation systems, they primarily focus on lower-order interactions. They often struggle to capture higher-order feature interactions.
\paragraph{3. Eliminating Useless Features:} Several existing methods have attempted to enhance feature interaction modelling in recommendation systems. AFM~\citep{AFM} introduces the concept of distinguishing between different feature interactions, but it falls short in eliminating cross-features. AIM~\citep{AFM} and AutoFIS~\citep{liu2020autofis} take a different approach by employing selection gating to prune irrelevant feature interactions. AutoFeature~\citep{khawar2020autofeature} introduces a NAS-based approach to identify essential feature interactions. Unfortunately, these methods often focus on specific aspects of feature interactions and fail to provide a comprehensive solution.
\section{Proposed Architecture: \trec}
\label{sec:method}
\trec{} utilizes a masked transformers-based approach (Figure~\ref{fig:transformer_interaction}) to handle data drift, minimize noise from irrelevant cross-features, and capture diverse patterns for higher-order interactions. Transformers are renowned for their ability to learn sequential correlations and encode word sequences using token embeddings and positional encoding.

In addition, \trec{} incorporates a Bottom MLP ($\mathrm{MLP_{bot}}$) to process dense inputs $\mathbf{x_{dense}}$, generating $\mathbf{y_{dense}}$ (Equation~\ref{eqn:1}). Sparse inputs $\mathbf{x_{sparse}}$ undergo an embedding lookup using the embedding table $\mathbf{E} \in \mathbb{R}^{M \times D}$, producing $\mathbf{y_{sparse}}$ (Equation~\ref{eqn:2}). Here, $M$ represents the number of items in the feature embedding $\mathbf{E}$, and $D$ denotes the sparse feature size.
\begin{align}
    \mathbf{y_{dense}} & = \mathrm{MLP_{bot}}(\mathbf{x_{dense}}) &  \label{eqn:1} \\
    \mathbf{y_{sparse}} & = \mathbf{x_{sparse}}\mathbf{E} & \mathbf{E} \in \mathbb{R}^{M \times D}   \label{eqn:2}
\end{align}

The outputs of dense and sparse features are concatenated to form the feature sequence $\mathbf{z_{0}} \in \mathbb{R}^{(N+1) \times D}$ (Equations~\ref{eqn:3} and \ref{eqn:4}). The feature sequence $\mathbf{z_{0}}$ captures joint embeddings of $N+1$ input features, where each feature is associated with a latent vector of size $D$.
\begin{align}
    \mathbf{z_{0}} & = [\mathrm{MLP_{bot}}(\mathbf{x_{dense}}) ; \mathbf{x_{sparse}^{1}}\mathbf{E^{1}} ; \ldots ; \mathbf{x_{sparse}^{N}}\mathbf{E^{N}}] & \label{eqn:3} \\  
    \mathbf{z_{0}} & = [\mathbf{y_{dense}} ; \mathbf{y_{sparse}^{1}} ; \ldots ; \mathbf{y_{sparse}^{N}}] \label{eqn:4}
\end{align}

This design empowers \trec{} to model feature interactions across multiple dimensions, addressing the challenge of capturing diverse patterns for higher-order cross-features. In deep learning-based recommender systems, the feature sequence $\mathbf{z_{0}}$ is either used directly or dot-product-based feature interactions are computed as $lower\;triangle(\;\mathbf{z_{0}} \times \mathbf{z_{0}^{T}}\;)$ to extract second-order cross-features.

\subsection{Masked Attention}

\begin{figure*}[t]
\centering
\subfloat[Masked Transformer-based Feature Interaction]{
\begin{minipage}[t]{0.45\textwidth}
\includegraphics[width=\textwidth]{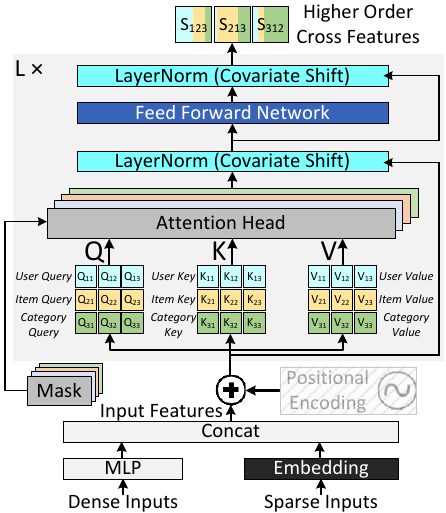}
\vskip -0.1in
\label{fig:transformer_interaction}
\end{minipage}}\hfill
\subfloat[Masked Multi-Head Self-Attention]{
\begin{minipage}[t]{0.42\textwidth}
\includegraphics[width=\textwidth]{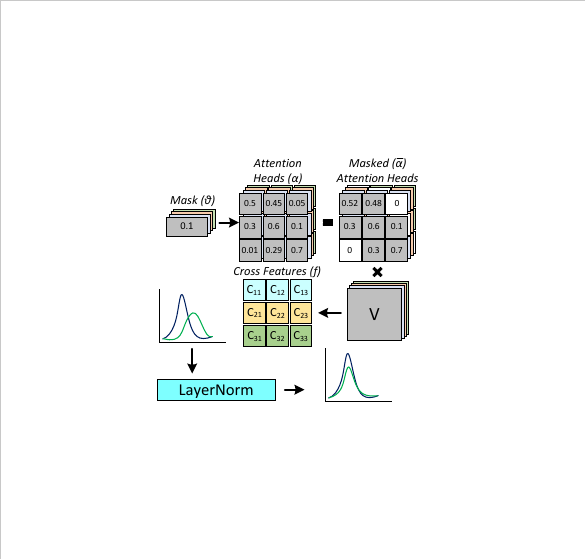}
\vskip -0.1in
\label{fig:masked_attention}
\end{minipage}}
\caption{(a) The masked transformer-based feature interaction enables efficient higher-order cross-features by using embedding tables and masking to eliminate irrelevant cross-features. LayerNorm further enhances cross-feature quality. (b) In the masked multi-head self-attention mechanism, scalar masks are assigned to each head to eliminate irrelevant cross-features, allowing for diverse cross-feature patterns in multiple subspaces. LayerNorm reduces internal covariate shift, leading to faster convergence and improved learning.}
\vspace{-0.1in}
\end{figure*}

The core component of \trec{} is the masked multi-head attention block, which enables explicit cross-feature creation in multiple subspaces while eliminating irrelevant ones. Figure~\ref{fig:masked_attention} visually demonstrates the masking process for removing irrelevant cross-features.

Each input feature $\mathbf{z_{i}}$ has query and key-value pairs learned in $\mathbf{h}$ subspaces, corresponding to the number of attention heads. This allows for capturing diverse cross-features in multiple subspaces. This is achieved through linear projection using matrices $\mathbf{W_{Q}^{h}}$, $\mathbf{W_{K}^{h}}$, and $\mathbf{W_{V}^{h} \in \mathbb{R}^{D \times D'}}$, where $\mathbf{D' = \frac{D}{h}}$. Specifically, the projected query, key, and value vectors for feature $\mathbf{i}$ in subspace $\mathbf{h}$ are denoted as $\mathbf{z_{qi}^{h} = z_{i}W_{Q}^{h}}$, $\mathbf{z_{ki}^{h} = z_{i}W_{K}^{h}}$, and $\mathbf{z_{vi}^{h} = z_{i}W_{V}^{h}}$, respectively. The correlation between feature $\mathbf{i}$ and feature $\mathbf{j}$ under a specific subspace $\mathbf{h}$ is represented by the attention head $\mathbf{\alpha_{i,j}^{h}}$ (Equation~\ref{eqn:5}), where $\mathbf{\langle \cdot \rangle}$ denotes the inner product.
\begin{align}
    \mathbf{\alpha_{i,j}^{h}} & = \frac{\exp \langle \mathbf{z_{qi}^{h}} \cdot \mathbf{z_{kj}^{h}} \rangle}{\sum_{m=1}^{N+1}{\exp \langle \mathbf{z_{qi}^{h}} \cdot \mathbf{z_{km}^{h}} \rangle}} &  \label{eqn:5}
\end{align}

\paragraph{Masking:} \trec{} uses a masking technique to eliminate irrelevant feature interactions. Each head $\mathbf{h}$ has a mask $\mathbf{\theta}$. The masked attention score $\mathbf{\overline{\langle z_{qi}^{h} \cdot z_{kj}^{h} \rangle}}$ is calculated as follows (Equation~\ref{eqn:6}):
\begin{align} 
\mathbf{\overline{\langle z_{qi}^{h} \cdot z_{kj}^{h} \rangle}}
& = 
\begin{cases}
    \mathbf{\langle z_{qi}^{h} \cdot z_{kj}^{h} \rangle}, & \text{if} \mathbf{\alpha_{i,j}^{h}} > \mathbf{\theta^{h}} \\ -\infty, & \text{otherwise}
\end{cases} & \label{eqn:6}
\end{align}

The recalculated attention head $\mathbf{\overline{\alpha_{i,j}^{h}}}$ (Equation~\ref{eqn:7}) is obtained by applying the softmax function to the masked attention scores:
\begin{align}
\mathbf{\overline{\alpha_{i,j}^{h}}} = \frac{\exp \overline{\langle \mathbf{z_{qi}^{h}} \cdot \mathbf{z_{kj}^{h}} \rangle}} {\sum_{m=1}^{N+1}{\exp \overline{\langle \mathbf{z_{qi}^{h}} \cdot \mathbf{z_{km}^{h}} \rangle}}} \label{eqn:7}
\end{align}

Finally, the cross-feature of feature $\mathbf{i}$ in subspace $\mathbf{h}$ is updated by combining the relevant feature attentions $\mathbf{\overline{\alpha_{i,j}^{h}}}$ with the corresponding values $\mathbf{z_{v}^{h}}$ (Equation~\ref{eqn:8}):
\begin{align}
\mathbf{f_{i}^{h}} & = \sum_{m=1}^{N+1}{\overline{\alpha_{i,m}^{h}} \cdot \mathbf{z_{vm}^{h}}} &   \label{eqn:8}
\end{align}

$\mathbf{\theta^{h}}$ is a head-specific mask that is fixed before training. In this work, a geometric sequence of decreasing values, such as $\frac{1}{10^{1}}, \frac{1}{10^{2}}, \ldots, \frac{1}{10^{h}}$, was used for the masks of multiple heads. This choice eliminates irrelevant features. While a single mask could be used for all heads, different masks per head allow for generalization. Appendix~\ref{subsec:mask_analysis} presents an ablation study on different mask values.

\paragraph{LayerNorm:} \trec{} utilizes Layer Normalization (LayerNorm) to normalize the output of the masked attention layer and feedforward network, ensuring stable and efficient training.

LayerNorm ($LN$), when applied to $\mathbf{f_{i}^{h}}$ of the masked attention layer, is defined using Equation~\ref{eqn:9}.
\begin{equation}
\mathbf{\overline{f_{i}^{h}}} = \mathrm{LN}(\mathbf{f_{i}^{h}}) = \frac{\mathbf{f_{i}^{h}} - \boldsymbol{\mu_{i}^{h}}}{\sqrt{\boldsymbol{\sigma_{i}^{2}} + \epsilon}} \odot \boldsymbol{\gamma^{h}} + \boldsymbol{\beta^{h}} \label{eqn:9}
\end{equation}

Here, $\boldsymbol{\mu_{i}^{h}}$ and $\boldsymbol{\sigma_{i}^{2}}$ are the mean and variance of $\mathbf{f_{i}^{h}}$ across feature dimensions, respectively. The $\odot$ operator represents element-wise multiplication. The learnable parameters $\boldsymbol{\gamma^{h}}$ and $\boldsymbol{\beta^{h}}$ scale and shift the normalized values. The term $\epsilon$ ensures numerical stability.

LayerNorm normalizes the output of the masked attention layer, $\mathbf{f_{i}^{h}}$, to have zero mean and unit variance across feature dimensions. This mitigates covariate shifts and provides a stable distribution for subsequent layers. The scale and shift parameters, $\boldsymbol{\gamma^{h}}$ and $\boldsymbol{\beta^{h}}$, enable the model to capture appropriate representations for the recommendation task.

LayerNorm also has a similar effect on the output of the feedforward network. By reducing reliance on the scale and distribution of the training dataset, it facilitates faster convergence, offers modest regularization, and improves the efficiency of parameter updates in the recommendation model.

\paragraph{Positional Embedding:} 
While positional embeddings are commonly used in language tasks to preserve word order, they are not utilized in \trec{} for recommender systems. Despite this, we evaluated 1-D positional embeddings to encode spatial information of features. The findings in Appendix~\ref{subsec:positional_emb} indicate that including positional embedding adversely affects the model's performance.

\trec{} employs masked multi-head attention ($MSA$) to learn explicit higher-order cross-features in multiple subspaces, and the feedforward network ($FFN$) handles implicit interactions. By stacking multiple \trec{} layers, up to $L$ (Equations~\ref{eqn:10} and~\ref{eqn:11}), higher-order interactions can be captured more effectively (see Appendix~\ref{subsubsec:encoder_layers}). It also improves performance for larger input feature sequences.
\begin{equation}
    \mathbf{z^{'}_{\ell}} = \mathrm{Masked \; MSA}(\mathrm{LN}(\mathbf{z_{\ell - 1}})) + \mathbf{z_{\ell - 1}}, \quad \ell = 1, \ldots, \mathbf{L} \label{eqn:10}
\end{equation}
\begin{equation}
    \mathbf{z_{\ell}} = \mathrm{FFN}(\mathrm{LN}(\mathbf{z^{'}_{\ell}})) + \mathbf{z_{\ell}}, \quad \ell = 1, \ldots, \mathbf{L} \label{eqn:11}
\end{equation}

The last layer's output is concatenated with processed dense inputs, resulting in $\mathbf{z}$ (Equation~\ref{eqn:12}).
\begin{equation}
    \mathbf{z} = [\mathbf{y_{dense}} ; \mathbf{z_{\ell}}] \label{eqn:12}
\end{equation}

The final click-through rate (CTR) is obtained by applying the top MLP ($\mathrm{MLP_{top}}$) to $\mathbf{z}$ (Equation~\ref{eqn:13}). Thus, the problem is modelled as binary classification using binary cross-entropy (BCE) loss to predict whether a user will click the target item.
\begin{equation}
    \mathrm{CTR} = \mathrm{MLP_{top}}(\mathbf{z}) \label{eqn:13}
\end{equation}

\subsection{Application to Sequential Recommendation Models}
\label{subsec:seq_rec}
For sequential recommendation models, we apply \trec{} by generating an embedding vector \( \mathbf{z}_t \) for each event \( \varepsilon_t \) using the non-sequential recommendation model. Here, \( \varepsilon_t \) represents the event at time step \( t \), such as a user click or purchase. These embeddings capture event characteristics, including explicit timing as a dense feature. The width of the last layer in the non-sequential model is adjusted to match the user-interaction vector's width, which is then fed into the sequential layer. This produces a sequence of embedding vectors \( Z \) (Equation~\ref{eqn:15}).
\begin{align}
    \mathbf{z}_{t} & = \mathbf{AdRec}(\varepsilon_{t}) &  t = 1, \ldots, \tau \label{eqn:14} \\
    Z & = [\mathbf{z}_{1}, \mathbf{z}_{2}, \ldots, \mathbf{z}_{\tau-1}] & \label{eqn:15} \\
    \mathbf{c} & = \mathbf{SequentialLayer}(\mathbf{z}_{\tau}, Z) \label{eqn:16} 
\end{align}

\trec{} generates an embedding \( \mathbf{z}_t \) for each event, capturing relevant information about the user-item interaction. These embeddings are used as input to the sequential layer, which considers the context and temporal order of earlier events to predict the next event \( \varepsilon_{\tau} \) at time step \( \tau \) (Equation~\ref{eqn:16}). By leveraging the \trec{} embeddings, the model effectively captures user-item interactions within the sequence, leading to improved recommendation performance.

\section{\trec{} Analysis}
\label{sec:trec_analysis}
To analyze the feature interaction of DLRM and \trec{}, we randomly sampled a test input from the real-world Taobao user behaviour dataset~\citep{alibaba}. This input consists of sequential user activity with a length of 21, a timestamp as a dense feature, and sparse features for user, item, and category. Notably, the ground truth of the sampled input indicates that it is a \textbf{negative sample}.

\begin{wrapfigure}{r}{0.5\linewidth}
\vskip -0.2in
  \centering
  \subfloat{
	\begin{minipage}[t]{0.405\linewidth}
	   \centering
	   \includegraphics[width=\textwidth]{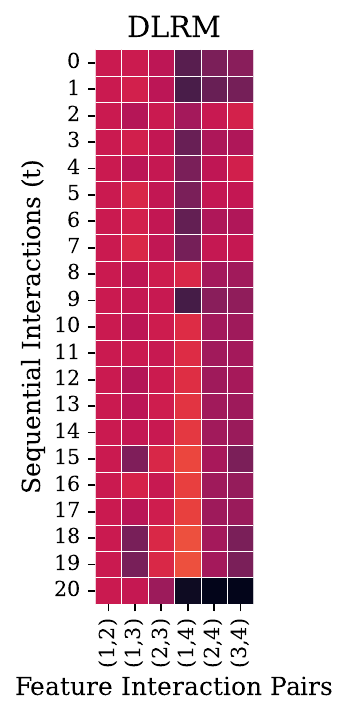}
	\end{minipage}}
  \subfloat{
	\begin{minipage}[t]{0.48\linewidth}
	   \centering
	   \includegraphics[width=\textwidth]{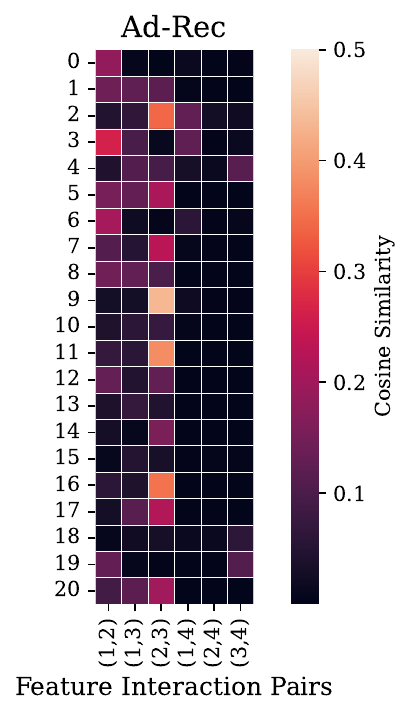}
	\end{minipage}}
\caption{Cosine similarity of feature interaction in the real-world Taobao dataset: one dense and three sparse features for a single user.}
\label{fig:cos_similarity}
\vspace{-3ex}
\end{wrapfigure}

Figure~\ref{fig:cos_similarity} presents the cosine similarity heat map, comparing the feature interaction of DLRM and \trec{}. In DLRM, the feature interaction is based on dot product calculations, while \trec{} employs masked attention. The features involved in the interaction include the user's activity timestamp (feature 1), user ID (feature 2), item ID (feature 3), and item category (feature 4). 

Examining the sequential interaction at timestamp $t=2$ in Figure~\ref{fig:cos_similarity}, we observe contrasting behaviour between DLRM and \trec{}. We observe that DLRM's dot product-based feature interaction provides close similarity across all pairs. Contrary to this, \trec{}'s masked attention-based feature interaction considers Query~($\mathbf{Q}$) and Key~($\mathbf{K}$) projections. Applying a mask with $\frac{1}{10^{3}}$ value reveals no features being masked (note that the mappings of attention heads can be found in Appendix~\ref{subsec:masked_attention_heads}).

The highest weight is assigned to feature pair $(2, 3)$, indicating a strong correlation between user ID and item ID. In contrast, other pairs such as $(1,2)$, $(1,4)$, $(2,4)$, and $(3,4)$ show weak alignment, implying a primarily \textbf{negative sample}. Unlike DLRM's dot product-based feature interaction, \trec{} captures this information.
\section{Experiments and Results}
\label{sec:experiments}

\subsection{Evaluation Setup}

\paragraph{Recommendation Models:} We train recommendation models with varying sizes to represent different classes of at-scale models~\citep{facebookreco:hpca, adnan2021highperformance, adnan2022heterogeneous}. The architecture of these models is presented in Table~\ref{table:non_seq_models}~\citep{facebook:ml, MTWnD}.

These models are selected based on their diverse characteristics and parameter sizes. The sparse parameter count ranges from 5.2M (RM4) to 266M (RM2), indicating different model complexity and capacity levels. It is worth noting that RM1 and RM2 are embedding-dominated models. On the other hand, RM3 is an average-sized model with a balanced mix of dense and sparse features.

To further explore the impact of sequential information, we consider models RM4, RM5, RM6, and RM7, which share a common base model for generating embedding vectors per event. However, they differ in the choice of the sequential layer to process the temporal sequence. RM4 employs a single time series layer (TSL), RM5 utilizes multi-head attention with eight heads, RM6 consists of 5 RNN layers, and RM7 employs a single transformer layer to predict the probability of the next event.

\begin{table*}[h!]
\centering
\caption{Recommendation Models Architecture and \trec{} Configuration (\# layers $=$ 1)}
\resizebox{1\textwidth}{!}{
\begin{tabular}{c | l | c c | c c c | c c | c c c | c c}
\hline
\multirow{3}{*}{\textbf{Model}} & \multirow{3}{*}{\textbf{Dataset}} &
\multicolumn{2}{c|}{\textbf{Features}} & \multicolumn{3}{c|}{\textbf{Parameters}} &  \multicolumn{2}{c|}{\textbf{Neural Network Configuration}} & \multicolumn{3}{c|}{\textbf{\trec Configuration}} & \multicolumn{2}{c}{\textbf{Sequential Layer}}
\\
\cline{3-14} &  & \multirow{2}{*}{\textbf{Dense}} & \multirow{2}{*}{\textbf{Sparse}} & \multirow{2}{*}{\textbf{Dense}} & \multirow{2}{*}{\textbf{Sparse}} & \textbf{Sparse} & \textbf{Bottom} & \textbf{Top} & \textbf{Num} & \textbf{Hidden} & \textbf{FFN} & \multirow{2}{*}{\textbf{Type}} & \textbf{Layers/} \\
 & & & & & & \textbf{Dim} & \textbf{MLP} & \textbf{MLP} & \textbf{Heads} & \textbf{Size} & \textbf{Config.} & & \textbf{Heads} \\
\hline
RM1 & Criteo Kaggle & 13 & 26 & 287.5k & 33.8M & 16 & 13-512-256-64-16 & 512-256-1 & 2 & 16 & 128 & \multirow{3}{*}{N/A} & \multirow{3}{*}{N/A} \\

RM2 & Criteo Terabyte & 13 & 26 & 549.1k & 266M & 64 & 13-512-256-64 & 512-512-256-1 & 8 & 64 & 512 & &\\

RM3 & Avazu & 1 & 21 & 281.4k & 9.3M & 16 & 1-512-256-64-16 & 512-256-1 & 2 & 16 & 128 & & \\
\hdashline
RM4 & \multirow{4}{*}{Taobao Alibaba} & \multirow{4}{*}{1} & \multirow{4}{*}{3} & \multirow{4}{*}{7.3k} & \multirow{4}{*}{5.1M} & \multirow{4}{*}{16} & \multirow{4}{*}{1-16} & \multirow{4}{*}{22-15-15} & \multirow{4}{*}{2} & \multirow{4}{*}{16} & \multirow{4}{*}{128} & TSL & 1 \\
RM5 & &  &  &  &  &  &  &  &  & & & MHA & 8 \\
RM6 & &  &  &  &  &  &  &  &  & & &  RNN & 5 \\
RM7 & &  &  &  &  &  &  &  &  & & &  Transformer & 1 \\
 \hline
\end{tabular}}
\label{table:non_seq_models}
\vspace{-2ex}
\end{table*}

\paragraph{Datasets:} We investigate \trec{}'s performance using various real-world datasets for both non-sequential and sequential recommendation tasks. For non-sequential tasks, we employ three datasets: Criteo Kaggle~\citep{criteokaggle}, Criteo Terabyte~\citep{criteoterabyte}, and Avazu~\citep{avazu}. The Criteo Kaggle dataset, derived from the Display Advertising Challenge, predicts click-through rates (CTR) and captures user preferences. Criteo Terabyte, the largest publicly available click-log dataset, is commonly used to train non-sequential DLRM models. The Avazu dataset, obtained from a CTR prediction competition on Kaggle, provides insights into users' ad click behaviours for mobile ads.

For sequential recommendation models, we turn to the Taobao User Behavior dataset~\citep{alibaba}, encompassing a vast array of 4 million items, 10,000 categories, and 1 million users. Within this dataset, each user's time series comprises triplets $(i, c, t)$, representing the interactions where a user $u$ engages with an item $i$ from category $c$ at time $t$. With an average of 21 sequential interaction events per user, the Taobao User Behavior dataset offers a rich environment to evaluate \trec{}.

\paragraph{Baselines:}
We compare \trec against four state-of-the-art techniques. DLRM~\citep{dlrm} and DeepFM~\citep{DeepFM} are $2^{nd}$ order feature interaction while AutoInt~\citep{autoint} and DCN-v2~\citep{dcnv2} are higher-order feature interaction techniques.

\paragraph{Training Details:} We conducted our experiments using PyTorch-1.9 and built upon the widely adopted Deep Learning Recommendation Model (DLRM)~\citep{dlrm} for non-sequential recommendations and Time-Based Sequence Model (TBSM)~\citep{tbsm} for the sequential counterpart. All models were implemented identically to ensure a fair comparison, with the feature interaction component being the only point of distinction.

For non-sequential models (RM1, RM2, and RM3), we employed the Stochastic Gradient Descent (SGD)~\citep{sgdtricks} optimizer. RM1 and RM3 were trained with a batch size of 128, while RM2 utilized a batch size of 1024. The learning rates were set to 0.01 for RM1, 0.1 for RM2, and 0.2 for RM3. In the case of sequential models (RM4, RM5, RM6, and RM7), we employed the Adagrad ~\citep{adagrad} optimizer with a learning rate of 0.05 and a batch size of 128. 

We performed five independent runs and presented the mean and standard deviation of the results. All models, except RM2, were trained on a single NVIDIA Tesla-V100 GPU. Due to the size of its embedding tables, RM2 was trained on 4 NVIDIA Tesla-V100 GPUs to ensure proper fitting.

\paragraph{Evaluation Metrics:} For evaluating the performance of our recommendation models, we use the Area Under Curve (AUC) metric, as established by the MLPerf community ~\citep{mlperf}. The AUC metric measures the probability that a randomly selected positive sample will be ranked higher than a randomly chosen negative sample. A higher AUC score indicates better performance. Each model is trained to achieve a specific target AUC score. Even a slight improvement in AUC, such as 0.001, is considered significant for click-through rate (CTR) prediction ~\citep{WnD, DeepFM, fignn}. We also track testing accuracy and BCE loss along with the AUC metric.

\begin{figure*}[t]
  \vspace{-3ex}
  \centering
  \subfloat{
	\begin{minipage}[t]{0.248\textwidth}
	   \centering
	   \includegraphics[width=\textwidth]{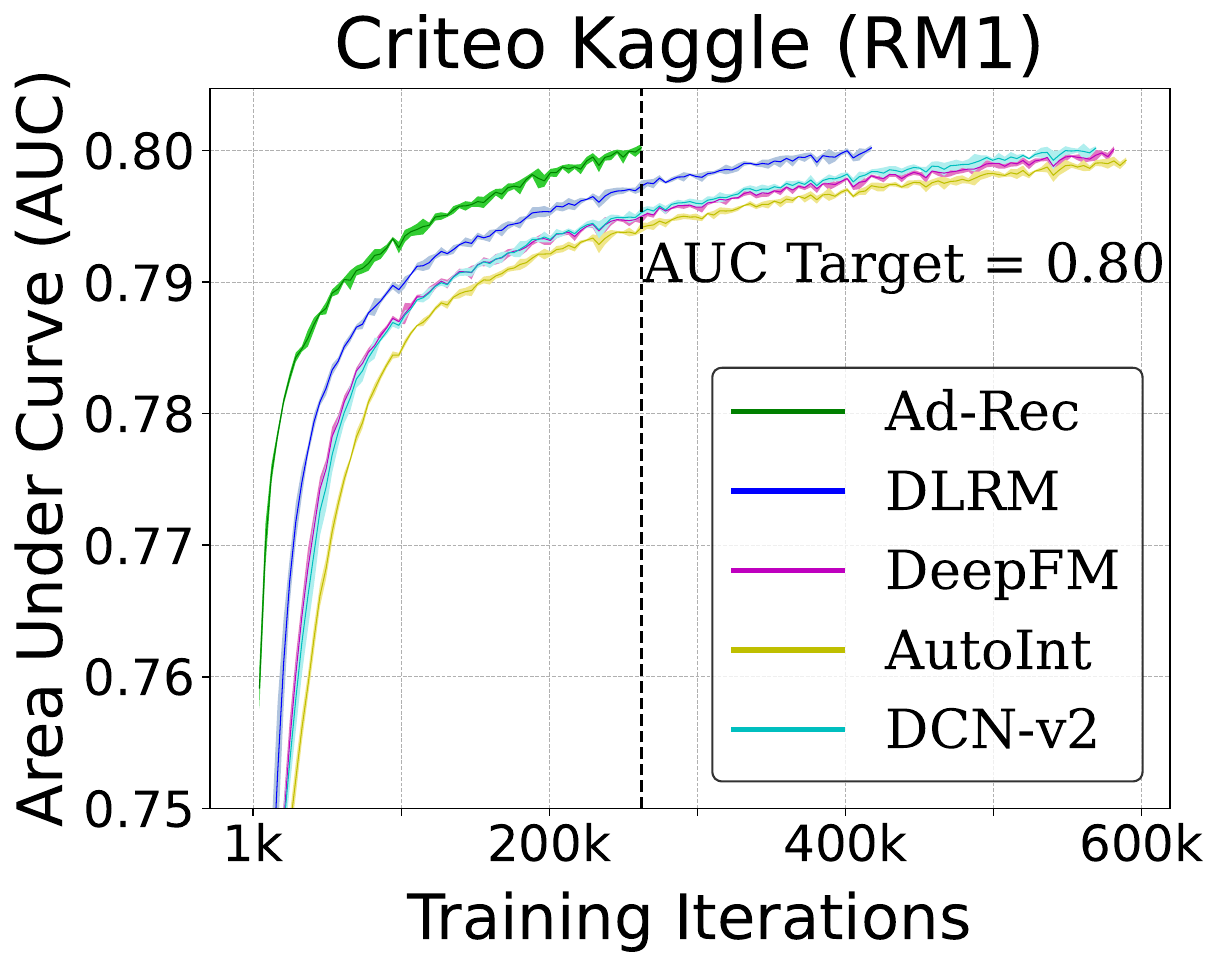}
	\end{minipage}}
  \subfloat{
	\begin{minipage}[t]{0.232\textwidth}
	   \centering
	   \includegraphics[width=\textwidth]{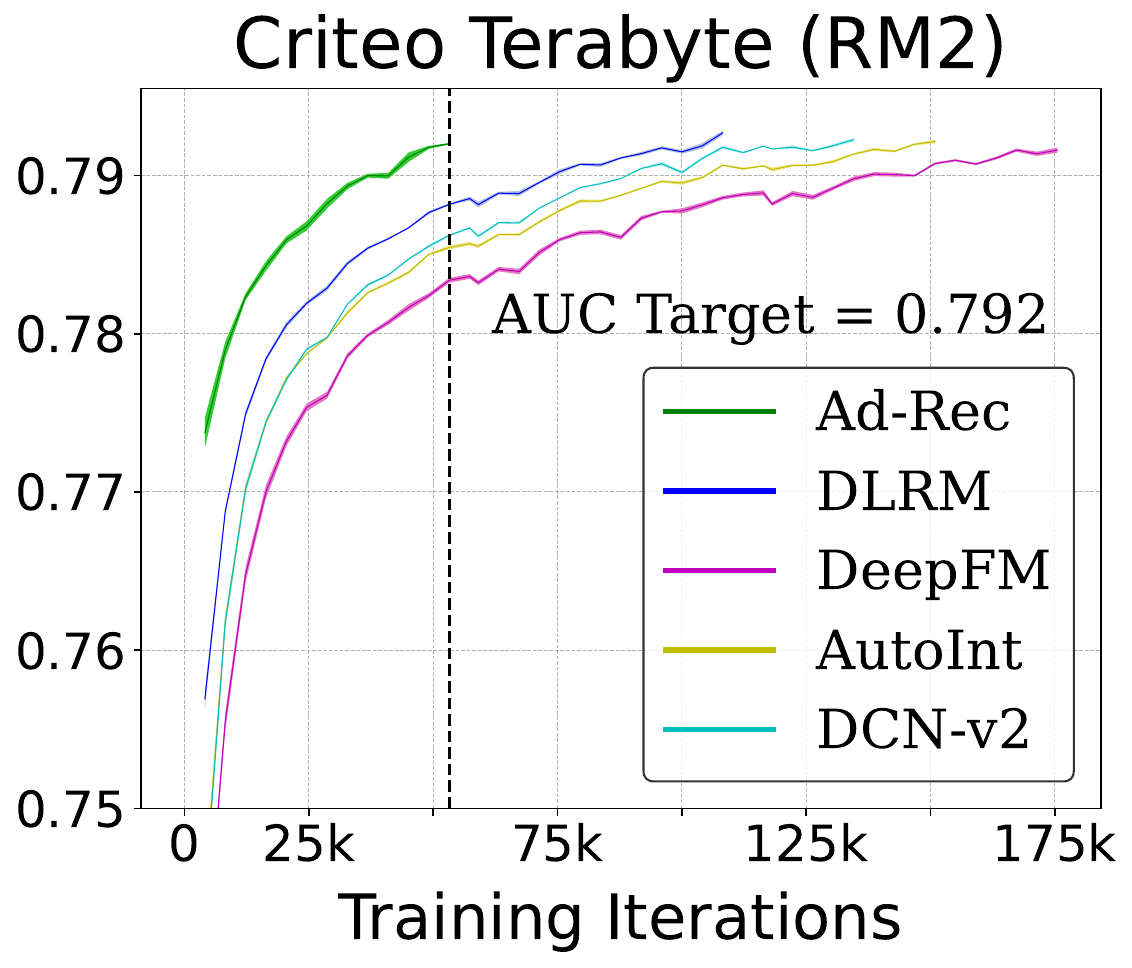}
	\end{minipage}}
  \subfloat{
	\begin{minipage}[t]{0.23\textwidth}
	   \centering
	   \includegraphics[width=1\textwidth]{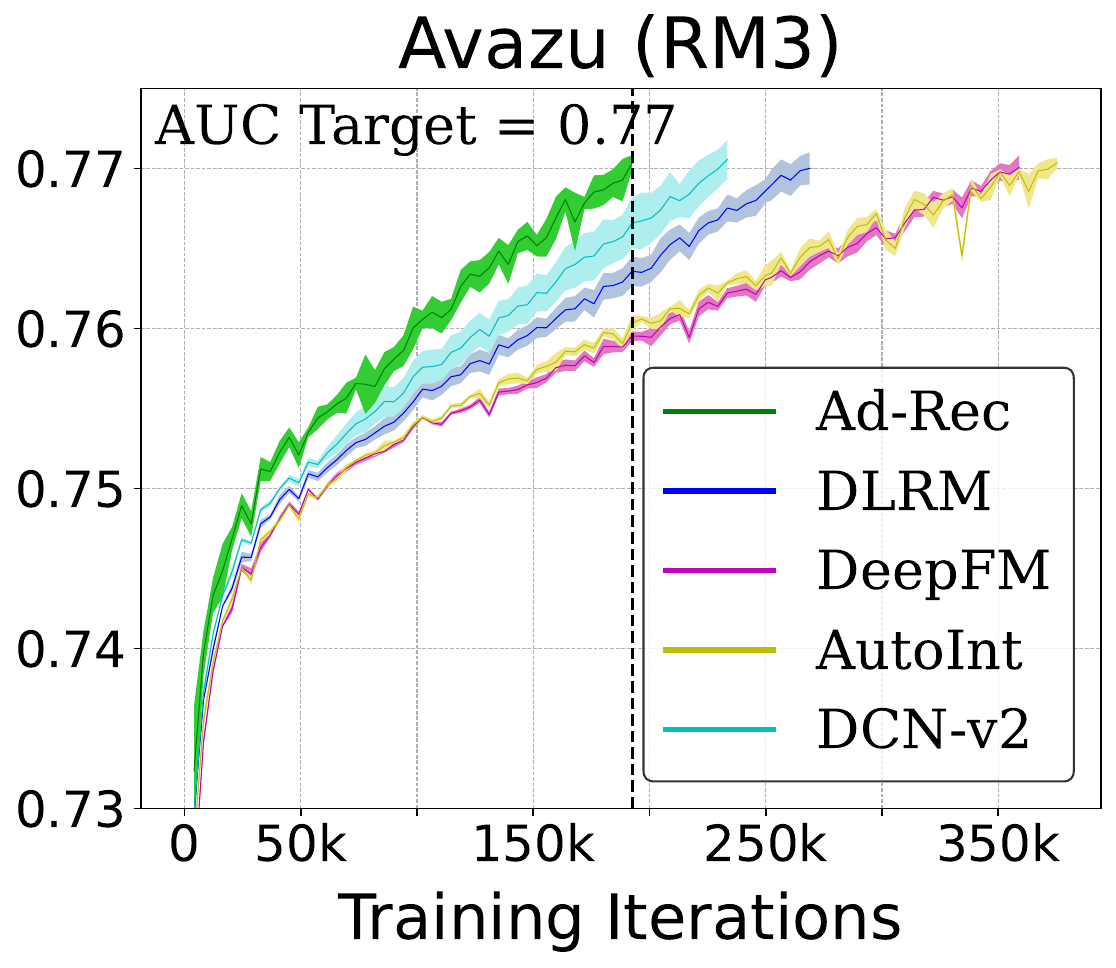}
	\end{minipage}}
 \subfloat{
	\begin{minipage}[t]{0.23\textwidth}
	   \centering
	   \includegraphics[width=1\textwidth]{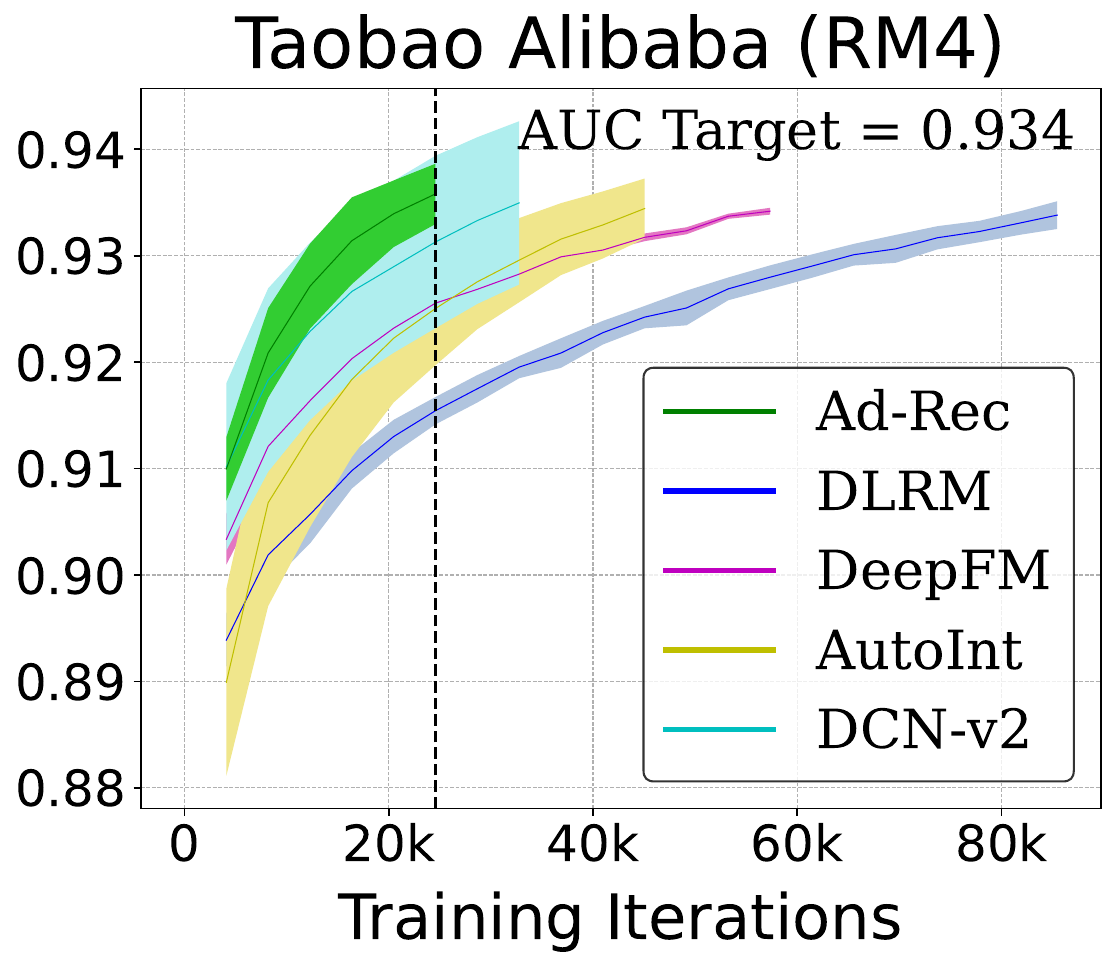}
	\end{minipage}}
\caption{Comparison of \trec{}'s convergence with baseline feature interaction techniques. The dotted vertical line represents the training iteration where \trec{} reaches the target AUC and stops training. On average, \trec{} achieves target AUC in \dlrmtrainiterimpr, \deepfmtrainiterimpr, \autointtrainiterimpr, and \dcntrainiterimpr lower iterations, as compared to the DLRM, DeepFM, AutoInt, and DCN-v2 baselines.}
\label{fig:conv_analysis}
\vspace{-3ex}
\end{figure*}

\subsection{Convergence Analysis}
\label{subsec:conv_analysis}

We compare the convergence of \trec{} with the baseline models (RM1, RM2, RM3, and RM4) by setting a target AUC for each model based on prior work~\cite{dlrm, tbsm}. The baseline models and \trec{} are trained with early stopping to achieve the target AUC. Figure~\ref{fig:conv_analysis} demonstrates that \trec{} achieves the target AUC in fewer training iterations, thanks to its masked attention-based feature interaction. The transformer encoder in \trec{} captures the relationships between input features, resulting in deeper feature representations and improved learning and prediction. On average, \trec{} achieves the target AUC in \dlrmtrainiterimpr, \deepfmtrainiterimpr, \autointtrainiterimpr, and \dcntrainiterimpr iterations, surpassing the DLRM, DeepFM, AutoInt, and DCN-v2 baselines.

Table~\ref{table:metric} compares evaluation metrics for a single training epoch. Across all models and datasets, \trec{} consistently outperforms the baselines regarding AUC. On average, \trec{} improves the AUC metric by \dlrmaucimpr, \deepfmaucimpr, \autointaucimpr, and \dcnaucimpr compared to the DLRM, DeepFM, AutoInt, and DCN-v2 baselines. Although higher-order feature interaction techniques may exhibit lower performance in certain models and datasets, \trec{} showcases superior generalization capabilities. It consistently outperforms other feature interaction techniques. For a comprehensive analysis of \trec{}'s hyperparameters, please refer to the ablation study in Appendix~\ref{subsec:rectra_hyperparameters}.

\begin{table*}[h!]
\vspace{-1ex}
\centering
\caption{Evaluation Metric Comparison with Single Epoch Training - Mean (stddev)}
\resizebox{1\textwidth}{!}{
\begin{tabular}{ l | c c c c c | c c c c c}
\hline
\multirow{2}{*}{\textbf{Model}} &  \multicolumn{5}{c|}{\textbf{AUC}} & \multicolumn{5}{c}{\textbf{BCE Loss}}
\\
 & \textbf{DLRM} & \textbf{DeepFM} & \textbf{AutoInt} & \textbf{DCN-v2} & \textbf{\trec} & \textbf{DLRM} & \textbf{DeepFM} & \textbf{AutoInt} & \textbf{DCN-v2} & \textbf{\trec} 
\\
\hline
RM1 & 0.798 (1.9e-4) & 0.796 (1.6e-4) & 0.795 (8.12e-5) & 0.796 (2.9e-4) & \textbf{0.801 (1e-4)}  & 0.459 (1.5e-3) & 0.461 (1.6e-3) & 0.461 (1.4e-3) & 0.460 (1.3e-3) & \textbf{0.455 (1.4e-3)}  \\

RM2 & 0.788 (1.8e-4) & 0.783 (1.5e-4) & 0.785 (1.5e-4) & 0.786 (8.29e-5) & \textbf{0.790 (1.8e-4)}  & 0.424 (1.4e-4) & 0.428 (1.5e-4) & 0.426 (7.25e-5) & 0.426 (5.14e-5) & \textbf{0.423 (9.4e-5)}  \\

RM3 & 0.768 (9.5e-4) & 0.763 (3.4e-3) & 0.763 (7.8e-4) & 0.772 (1e-3) & \textbf{0.775 (3.2e-4)}  & 0.386 (3.8e-4) & 0.390 (2.6e-4) & 0.390 (3.6e-4) & 0.384 (5e-4) & \textbf{0.382 (1.2e-4)}  \\

RM4 & 0.933 (1.3e-4) & 0.939 (6.6e-4) & 0.942 (1.5e-3) & 0.947 (5.8e-3) & \textbf{0.949 (1.9e-3)} & 0.267 (3.4e-3) & 0.257 (1.3e-3) & 0.254 (3.8e-3) & 0.237 (1.5e-2) & \textbf{0.232 (8.3e-3)}  \\

RM5 & 0.869 (4.2e-4) & 0.881 (1.1e-4) & 0.893 (5.5e-3) & 0.894 (1.5e-3) & \textbf{0.895 (2.1e-3)} & 0.378 (1.8e-4) & 0.360 (6.3e-3) & 0.363 (1.9e-3) & 0.361 (3.2e-3) & \textbf{0.361 (2e-3)}  \\

RM6 & 0.840 (7.1e-3) & 0.850 (1e-3) & 0.849 (2.8e-3) & \textbf{0.855 (1.3e-3)} & 0.852 (2.6e-3) & 0.385 (1.4e-3) & 0.388 (3.7e-3) & 0.388 (7.2e-3) & \textbf{0.375 (1.1e-3)} & 0.380 (4.2e-3)  \\

RM7 & 0.920 (1.8e-3) & 0.934 (1.3e-3) & 0.932 (6e-3) & \textbf{0.940 (5.2e-3)} & \textbf{0.940 (1.8e-3)} & 0.299 (1.1e-3) & 0.269 (2.8e-3) & 0.279 (1.2e-3) & 0.255 (1.3e-3) & \textbf{0.254 (5.6e-3)}  \\
\hline
\end{tabular}}
\label{table:metric}
\vspace{-1ex}
\end{table*}

\paragraph{Performance Comparison:}

\begin{wrapfigure}{r}{0.55\linewidth}
\begin{center}
\vskip -0.25in
\centerline{\includegraphics[width=1\linewidth]{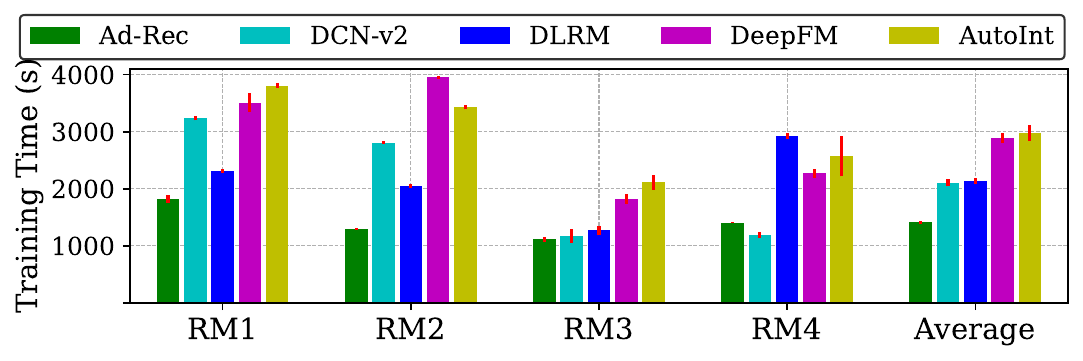}}
\caption{Absolute training time over multiple runs. \trec converges to target AUC in less time, which translates to a speedup of \dlrmtrainspeedup, \deepfmtrainspeedup, \autointtrainspeedup and \dcntrainspeedup over DLRM, DeepFM, AutoInt, and DCN-v2 baselines.}
\label{fig:speedup}
\end{center}
\vskip -0.3in
\end{wrapfigure}

\trec{}'s transformer-based feature interaction is computationally expensive compared to baselines. The wall clock time is measured to compare the runtime of training the recommendation model. As Figure~\ref{fig:speedup} shows, as expected, a single training iteration of \trec-based training takes more time. Still, it converges in less number of training iterations that provides a speedup of \dlrmtrainspeedup, \deepfmtrainspeedup, \autointtrainspeedup and \dcntrainspeedup over DLRM, DeepFM, AutoInt, and DCN-v2 baselines.

\begin{figure}[h!]
\begin{center}
\centerline{\includegraphics[width=0.65\columnwidth]{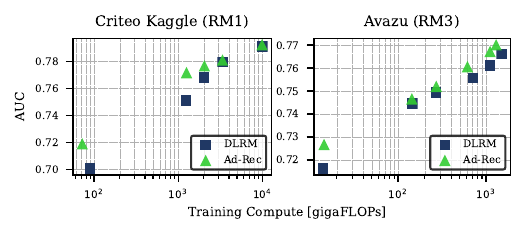}}
\caption{Model quality versus computational cost for different models. \rectra outperforms state-of-the-art DLRM with the same computational budget.}
\label{fig:compute}
\end{center}
\end{figure}

\paragraph{Computational Cost Analysis:} To compare the computational cost of \trec{}-based feature interaction, we trained DLRM and \trec{} using a fixed computational budget and compared the training quality metric (AUC). Figure~\ref{fig:compute} shows that \rectra dominates DLRM on this performance-compute trade-off. Similar trends are observed for other remaining models (RM2 and RM4).

\subsection{Ablation Study for Covariate Shift}

\begin{wraptable}{r}{0.6\textwidth}
\vspace{-0.1in}
\caption{Ablation study comparing the impact of LayerNorm ($LN$) in \trec to handle covariate shift.}
    \resizebox{0.6\columnwidth}{!}{
    \begin{tabular}{l|c c|c c}
         \hline
         \multirow{2}{*}{\textbf{Model}} &  \multicolumn{2}{c|}{\textbf{AUC}} & \multicolumn{2}{c}{\textbf{BCE Loss}} \\
          & \textbf{\trec w/ LN} & \textbf{\trec w/o LN} & \textbf{\trec w/ LN} & \textbf{\trec w/o LN} \\
         \hline
         RM1 & \textbf{0.801 (1e-4)} & 0.795 (2.4e-4) & \textbf{0.455 (1.4e-3)} & 0.460 (1.4e-3)\\
         RM2 & \textbf{0.790 (1.8e-4)} & 0.785 (2.2e-4) & \textbf{0.423 (9.4e-5)} & 0.426 (1.4e-4)\\
         RM3 & \textbf{0.775 (3.2e-4)} & 0.771 (1.8e-4) & \textbf{0.382 (1.2e-4)} & 0.385 (1.4e-4)\\
         RM4 & \textbf{0.949 (1.9e-3)} & 0.944 (2.8e-3) & \textbf{0.232 (8.3e-3)} & 0.247 (7.4e-3)\\
         RM5 & \textbf{0.895 (2.1e-3)} & 0.890 (2.8e-3) & \textbf{0.361 (2e-3)} & 0.366 (5.9e-3)\\
         RM6 & \textbf{0.852 (2.6e-3)} & 0.823 (7.3e-3) & \textbf{0.400 (4.2e-4)} & 0.427 (6.3e-3)\\
         RM7 & \textbf{0.940 (1.8e-3)} & 0.930 (2e-3) & \textbf{0.272 (5.6e-3)} & 0.284 (4.8e-3)\\
         \hline
    \end{tabular}}
    \label{tab:ablation_norm}
\end{wraptable}
We evaluate the importance of LayerNorm in \trec{} with an ablation study that removes the LayerNorm layer while keeping the rest of the architecture unchanged. Table~\ref{tab:ablation_norm} shows a decrease in the AUC metric and an increase in the BCE Loss when LayerNorm is omitted. Thus LayerNorm is critical to handling data distribution drift and promoting faster and more stable convergence in \trec{}.

\subsection{Scaling the Recommendation Models}
\begin{figure*}[h!]
  \vspace{-1.5ex}
  \centering
  \subfloat{
	\begin{minipage}[t]{0.25\textwidth}
	   \centering
\includegraphics[width=\textwidth]{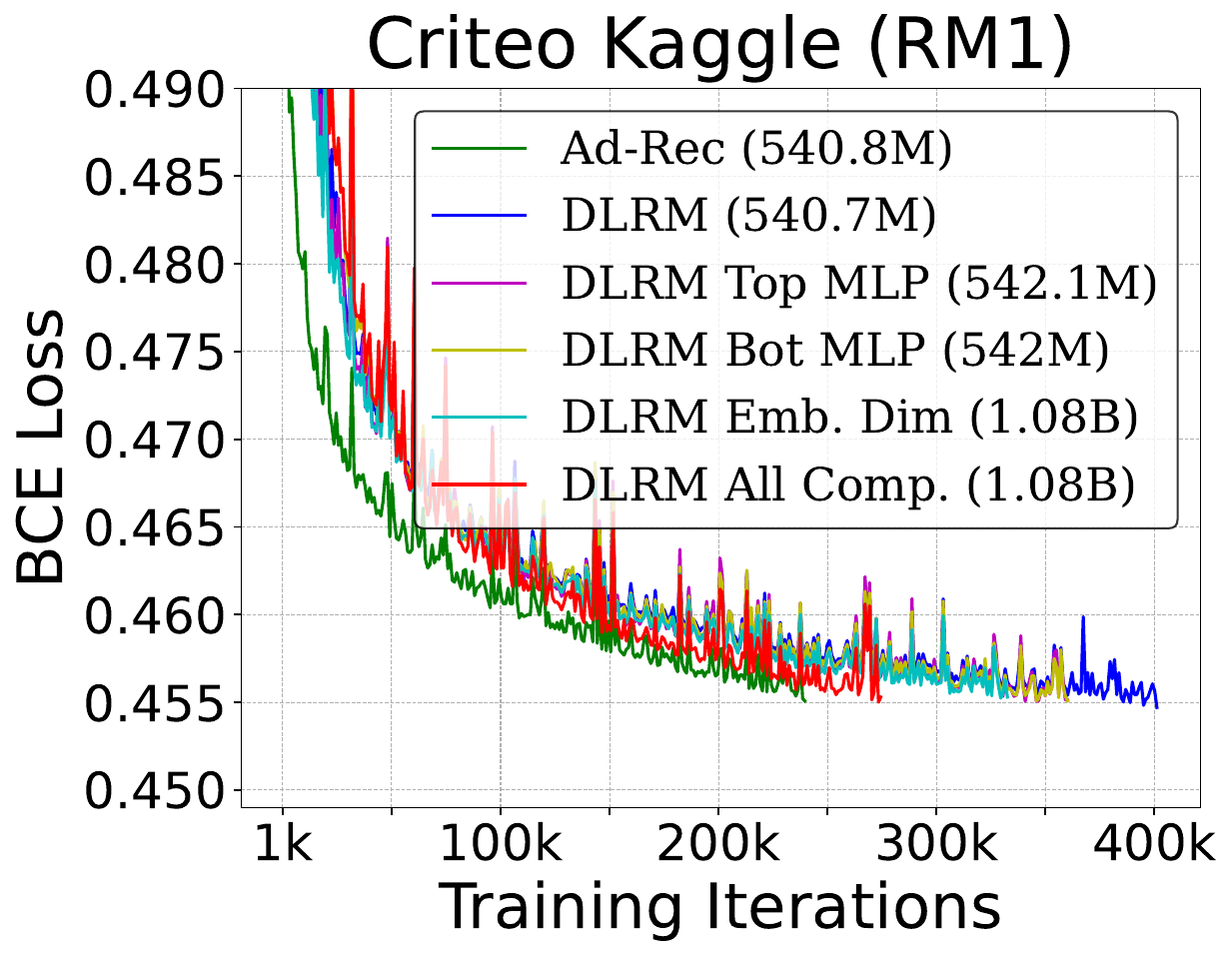}
	\end{minipage}}
  \subfloat{
	\begin{minipage}[t]{0.232\textwidth}
	   \centering
	   \includegraphics[width=\textwidth]{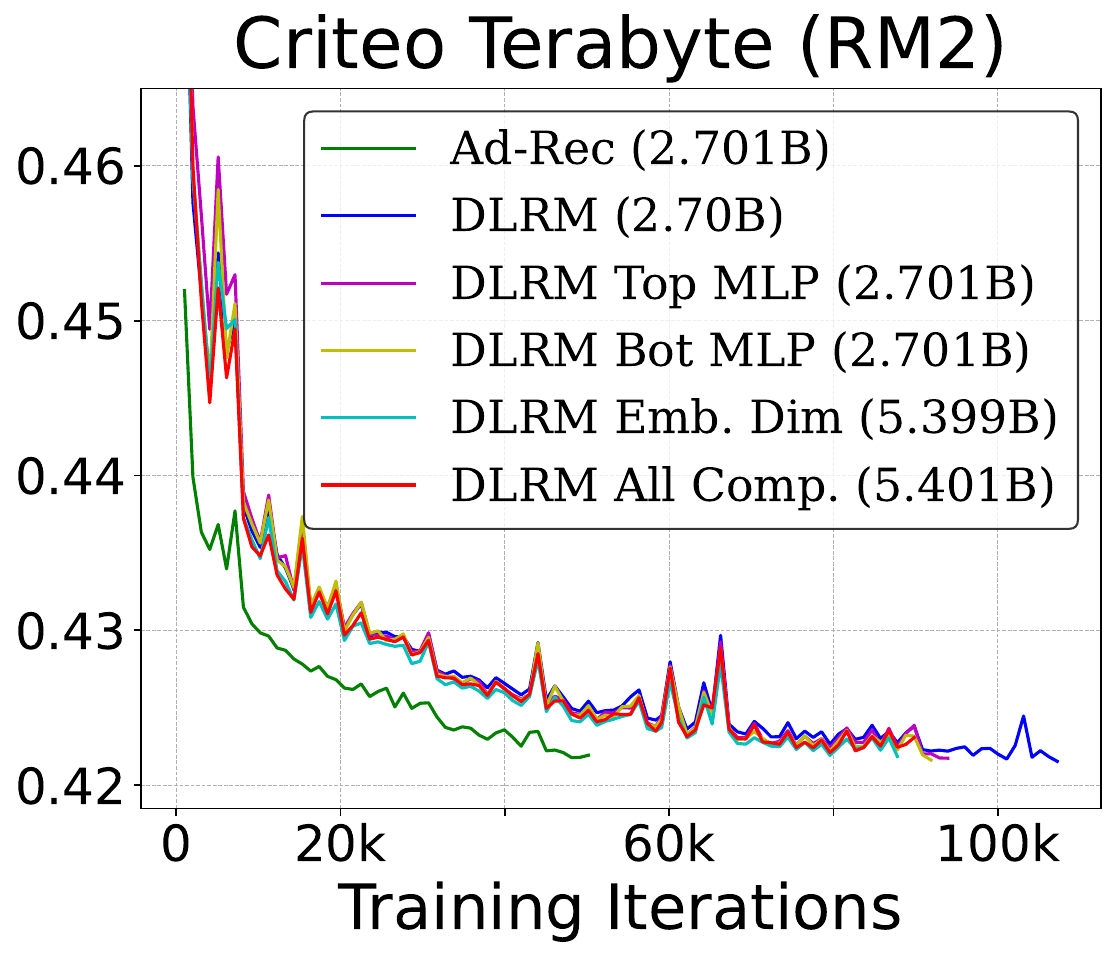}
	\end{minipage}}
 \begin{minipage}[t]{0.24\textwidth}
	   \centering
	   \includegraphics[width=\textwidth]{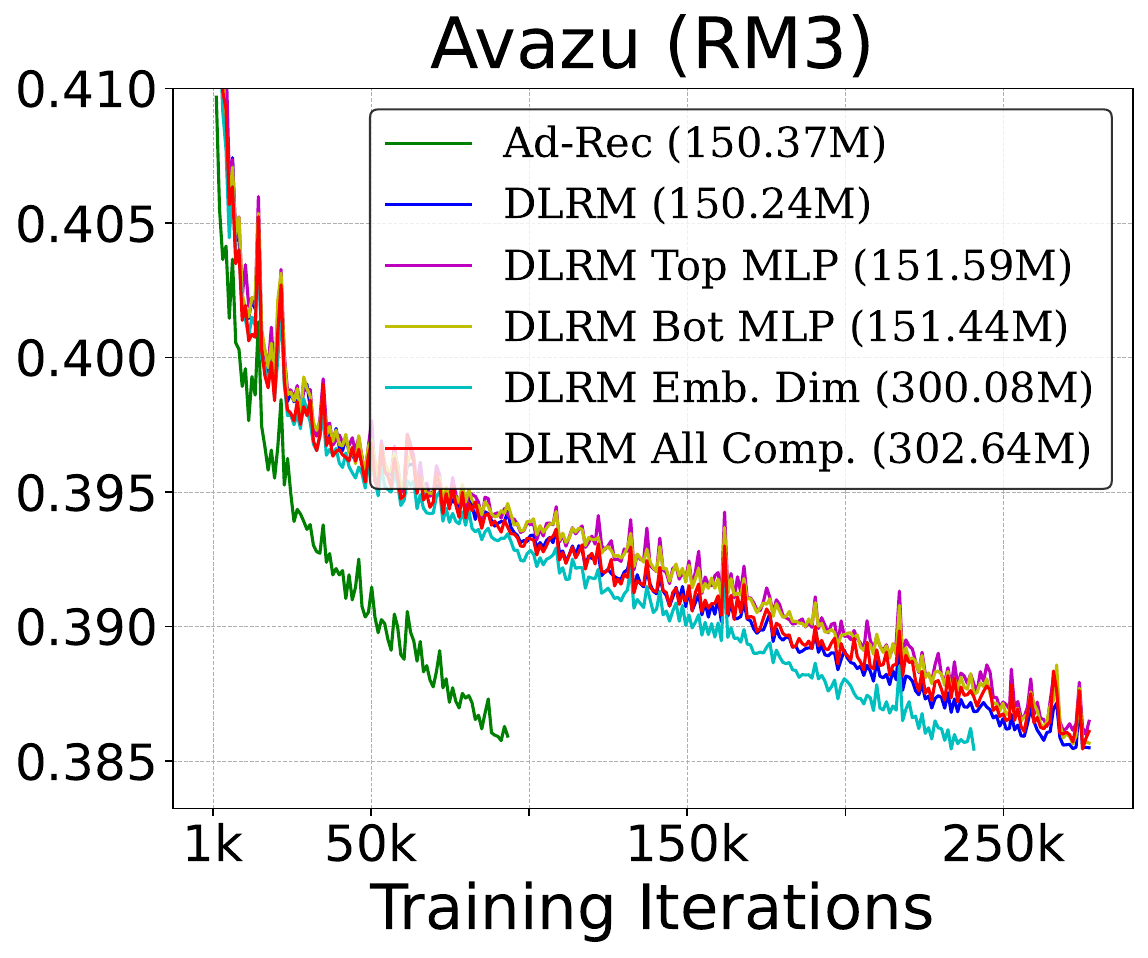}
	\end{minipage}
  \subfloat{
	\begin{minipage}[t]{0.235\textwidth}
	   \centering
	   \includegraphics[width=\textwidth]{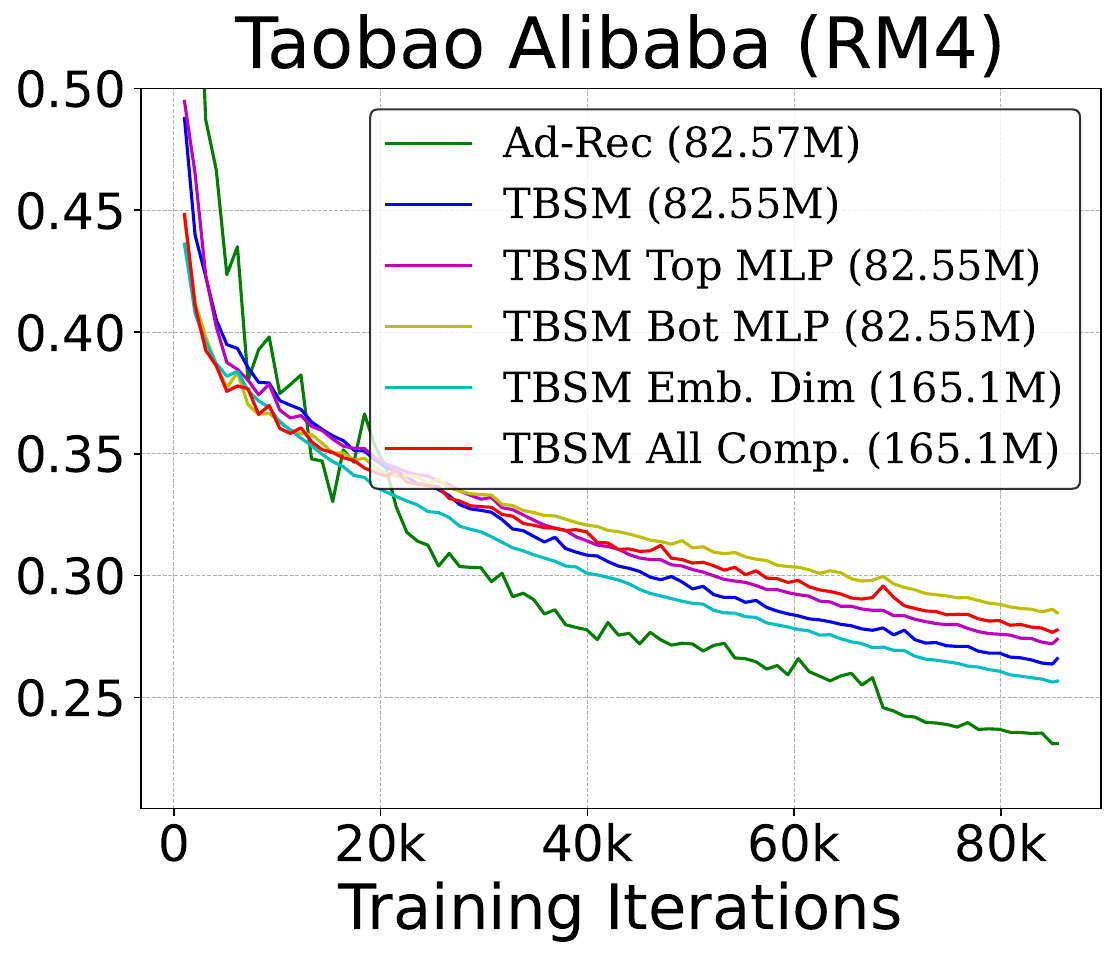}
	\end{minipage}}
\caption{Scaling laws of recommendation models were studied to understand the impact of component scaling on model quality. Despite a small training dataset, \trec{} outperformed scaled models.}
\label{fig:scaling_laws}
\vskip -0.1in
\end{figure*}

We conducted scaling ablation studies on DLRM-style recommendation models. Scaling the embedding dimension had the most significant improvement, while scaling the model size had minimal effect on model quality. More details can be found in Appendix~\ref{subsec:scaling_laws}. In the ablation studies with \trec{}, we evaluated model quality using masked-attention-based feature interaction while keeping the model size and computational budget fixed. Figure~\ref{fig:scaling_laws} shows that \trec{} consistently outperformed other models across parameters, such as embeddings, embedding dimension, and neural network size. 

\section{Limitations and Future Work}
\label{sec:limitations}
While \trec{} exhibits promising results, two key areas warrant further investigation. First, the masked attention-based feature interaction in \trec{} enhances model performance but may compromise interpretability compared to other techniques. Future work can focus on developing approaches to improve inter-operability. Second, recent higher-order feature interaction techniques, including \trec{} and AutoInt~\citep{autoint}, exhibit increased inference time due to their higher computational complexity for attention-based operations. Meeting strict Service Level Agreements (SLAs) becomes challenging with such higher-order techniques. Future research could explore leveraging the learned features from \trec{} and applying less computationally expensive feature interaction techniques.
\section{Conclusion}
\label{sec:conclusion}
Recommendation models are pivotal in improving recommendation quality and user experience, attracting considerable attention from industry and academia. In this study, we tackled the challenges of covariate shifts and sought to enhance cross-feature learning while accounting for data drift. Our findings underscored the significance of normalizing explicit cross-features and eliminating noisy ones to enhance recommendations in unfamiliar data distributions. To address these challenges, we introduced \trec{}, a masked-attention-based feature interaction technique. Through rigorous experimentation, \trec{} consistently outperformed state-of-the-art baselines, exhibiting superior quality and convergence speed. It achieved higher AUC scores and accelerated the training process, delivering compelling results across commercial models and publicly available datasets.





\bibliography{main}
\bibliographystyle{plainnat}

\newpage
\appendix
\onecolumn
\section{Appendix}

\subsection{High-Level Overview: Deep Learning-based Recommendations with \trec{}}
\label{subsec:background}
Figure~\ref{fig:recmodel} illustrates the model architecture of \trec{}, a non-sequential deep learning-based recommendation model inspired by DLRM~\citep{dlrm}. The model takes two types of inputs: dense and sparse features. Dense inputs consist of continuous features such as the user's age or the time of day. In contrast, sparse inputs include categorical features like the user's location or liked videos. 

Multi-layer perceptrons (MLPs) are employed to process the dense inputs, while large embedding tables handle the sparse inputs, each representing a specific categorical feature. These embeddings, along with the dense features, are then passed through a feature interaction layer, enabling the generation of cross-features that capture complex relationships between different features. Next, the cross-features and dense features are fed into an MLP layer, which predicts the click-through rate (CTR). The CTR indicates the likelihood of a user clicking on an item, serving as a key metric for recommendation models. By leveraging the power of deep learning and effective feature interaction, \trec{} enhances the accuracy and performance of recommendation systems.

Figure~\ref{fig:seqrecomodel} provides an overview of sequential recommendation models within the context of \trec{}. These models explicitly incorporate the temporal aspect of user-item interactions through a set of events denoted as $\varepsilon$. Each event $\varepsilon$ represents a user $u$ interacting with an item $i$ at a specific time $t$. 

User behaviour is captured as a sequence of events $\varepsilon$ spanning multiple time steps, denoted as $t = \{1, \ldots, \tau-1\}$. The sequential recommendation model is then trained to predict the next event at time step $\tau$, based on the previous event sequence. This shift towards sequential modelling transforms the organization of the training dataset, as it now represents a sequence of user activities rather than a simple collection of user interactions with features, as seen in non-sequential recommendation models. By incorporating temporal dynamics, \trec{} enables more accurate and personalized recommendations in dynamic user environments.

\subsection{Multihead Self-Attention}
\label{subsec:MSA}
The standard self-attention mechanism, commonly used in Natural Language Processing (NLP) tasks, forms a fundamental building block. Given an input sequence $\mathbf{z} \in \mathbb{R}^{N \times D}$, it computes a weighted sum over all values $\mathbf{v}$ in the sequence. The attention weights $\mathbf{\mathrm{A_{ij}}}$ reflect the pairwise similarity between query $\mathbf{q_{i}}$ and key $\mathbf{v_{j}}$ representations of the input tokens.

\begin{flalign}
    \hspace{0.6in}
    \mathbf{q, k, v} & = \mathbf{zU}_{qkv} & \mathbf{U}_{qkv} \in \mathbb{R}^{D \times 3D} & \quad \quad \label{eqnapp:17}\\
     \hspace{0.6in}
     \mathbf{A} & = \mathrm{softmax}(\frac{\mathbf{qk^{T}}}{\sqrt{\mathbf{D}}}) & \mathbf{A} \in \mathbb{R}^{N \times N} & \quad \quad \label{eqnapp:18}\\
     \mathrm{SA} & = \mathbf{Av} & & \label{eqnapp:19}
\end{flalign}

Multihead self-attention (MSA) extends the self-attention mechanism by performing $H$ parallel self-attention operations, where each self-attention is referred to as a ``head''. The parameter $H$ is a hyper-parameter that determines the number of heads. To ensure constant computation and parameter count with respect to $H$, the dimension $D$ (Eq.~\ref{eqnapp:17}) is set to $\frac{D}{H}$.

\begin{flalign}
    \hspace{0.6in}
    \mathrm{MSA}(\mathbf{z}) & = [\mathrm{SA}_{1}(\mathbf{z}) \;; \mathrm{SA}_{2}(\mathbf{z}) \;; \ldots \;; \mathrm{SA}_{n}(\mathbf{z})] \mathbf{U}_{msa} & \mathbf{U}_{msa} \in \mathbb{R}^{HD \times D} & \quad \quad \label{eqnapp:20}
\end{flalign}

\subsection{Mask Analysis}
\label{subsec:mask_analysis}
In our ablation study, we examined the impact of masking and the choice between a fixed mask value or different mask values for each attention head. Three scenarios were considered: no mask, a fixed mask value for all heads, and different mask values for each head. Table~\ref{table:mask_analysis} illustrates the advantages of employing distinct mask values for each head, highlighting the superiority over using a fixed mask value or no masking.

\begin{table*}[h!]
\vspace{-1ex}
\centering
\caption{Ablation Study. Mask $(\mathbf{\theta})$ Analysis with Single Epoch Training - AUC Mean (std deviation). The `No Mask' entry indicates a scenario containing no masking. The numbers in other entries show fixed mask values across all heads, and \trec{} employs different mask values across each head.}
\resizebox{1\textwidth}{!}{
\begin{tabular}{ l | c c c c c c c}
\hline
\multirow{2}{*}{\textbf{Model}} &  \multicolumn{7}{c}{\textbf{Mask Value $(\theta)$}}
\\
 & \textbf{No Mask} & \textbf{0.1} & \textbf{0.01} & \textbf{0.05} & \textbf{0.001} & \textbf{0.005} & \textbf{\trec} 
\\
\hline
RM1 & 0.801095 (3.17e-4) & 0.801269 (3.74e-4) & 0.801357 (2.2e-4) & 0.801055 (4.33e-4) & 0.801112 (4.82e-4) & 0.801199 (1.14e-4) & \textbf{0.80153 (1.09e-4)} \\

RM2 & 0.790164 (2.44e-4) & 0.790222 (1.84e-4) & 0.790214 (1.92e-4) & 0.790176 (8.07e-5) & 0.790185 (4.56e-4) & 0.790179 (1.71e-4) & \textbf{0.790224 (1.56e-4)} \\

RM3 & 0.775091 (4.85e-4) & 0.775081 (7.9e-4) & 0.775044 (4.08e-4) & 0.775344 (9.2e-4) & 0.775637 (2.1e-4) & 0.77587 (3.21e-4) & \textbf{0.775971 (4.3e-4)} \\

RM4 & 0.9453581 (2.88e-3) & 0.9421871 (6.46e-3) & 0.9465051 (2.57e-3) & 0.9430625 (3.67e-3) & 0.9434243 (4.54e-3) & 0.9406334 (1.45e-3) & \textbf{0.9494345 (1.9e-3)} \\

RM5 & 0.8940437 (2.87e-4) & 0.8938943 (2.93e-4) & 0.89249 (2.04e-3) & 0.8866167 (9.16e-3) & 0.8882817 (5.43e-3) & 0.8892716 (4.01e-3) & \textbf{0.8955358 (7.65e-3)} \\

RM6 & 0.8500284 (1.73e-3) & 0.840935 (8.25e-3) & 0.8453409 (6.09e-3) & 0.8424109 (6.67e-3) & 0.8491047 (5.64e-3) & 0.8374637 (4.19e-3) & \textbf{0.8520705 (2.61e-3)} \\

RM7 & 0.9323605 (1.45e-3) & 0.9319719 (5.68e-3) & 0.9326281 (1.05e-3) & 0.9337075 (1.84e-3) & 0.9317426 (8.62e-3) & 0.9327804 (1.45e-3) & \textbf{0.9404412 (1.82e-3)} \\
\hline
\end{tabular}}
\label{table:mask_analysis}
\vspace{-1ex}
\end{table*}

Models trained with masking generally demonstrate improved AUC compared to models without masking, as masking effectively eliminates irrelevant cross-features that can degrade prediction quality. However, determining the optimal mask threshold presents a challenge. Different models require different mask thresholds ($\theta$), and selecting an unsuitable mask can harm prediction quality by eliminating important features. Prior approaches~\citep{liu2020autofis, khawar2020autofeature, AIM} tackle this issue by training models specifically to learn such features, but this approach incurs computational costs and lacks generalizability across models or even the same model with different sizes. In contrast, \trec{} addresses this challenge by utilizing different mask values for each attention head, facilitating better generalization. If an important feature is eliminated in one head, it can be compensated for by other attention heads, resulting in a higher-quality model.

\subsection{Positional Embedding}
\label{subsec:positional_emb}

In our sensitivity study, we explored different ways of encoding the spatial information of features using positional embeddings inspired by NLP models. We considered two cases:
\begin{itemize}
    \item \textbf{No positional information:} This case involved using only feature embeddings and providing them as-is to the transformer encoder. This approach was the default across all other experiments in the paper.
    \item \textbf{1-dimensional positional embedding:} We treated the input features as a sequence of features, assigning each sparse feature and dense feature vector a position based on the embedding table position for language models. We added position embeddings to the feature inputs just before feeding them to the Transformer encoder. The dimension of the position embedding was kept similar to the sparse feature dimension, and the number of position embeddings was equal to the number of features.
\end{itemize}

Equation~\ref{eqnapp:21} demonstrates the incorporation of positional embeddings into the input features. The feature inputs and their respective embeddings were summed with the position embeddings.

\begin{flalign}
    \mathbf{z_{0}} & = [\mathrm{MLP}(\mathbf{x_{dense}}) ; \mathbf{x_{sparse}^{1}}\mathbf{E^{1}} ; \; . \; . \; . \; . \; ; \; \mathbf{x_{sparse}^{N}}\mathbf{E^{N}}\;] + \mathbf{E_{pos}} & \mathbf{E} \in \mathbb{R}^{M \times D} , \; \mathbf{E_{pos}} \in \mathbb{R}^{(N+1) \times D} \label{eqnapp:21}
\end{flalign}

Table~\ref{table:pos_emb} presents the evaluation results comparing the models with and without positional embeddings. Our hypothesis was that since the Multi-Head Attention block exhibits permutation-equivariance, the position of a feature in the input sequence does not encode useful information. Therefore, the models that directly input the raw features to the masked transformer encoder, facilitating higher-order feature interaction, performed the best. Incorporating 1-dimensional positional embeddings led to a degradation in model performance, even worse than the DLRM baseline. Notably, in certain models like RM3, incorporating 1-dimensional positional embeddings prevented the model from converging.

\begin{table}[h!]
\centering
\caption{Results of ablation study on positional embeddings with single epoch training.}
\resizebox{0.8\columnwidth}{!}{
\begin{tabular}{ l | c c | c c | c c}
\hline
\multirow{2}{*}{\textbf{Model}} &  \multicolumn{2}{c|}{\textbf{AUC}} & \multicolumn{2}{c|}{\textbf{Test Accuracy (\%)}} & \multicolumn{2}{c}{\textbf{BCE Loss}} 
\\
 & \textbf{No Pos. Emb.} & \textbf{1-D Pos. Emb.} & \textbf{No Pos. Emb.} & \textbf{1-D Pos. Emb.} & \textbf{No Pos. Emb.} & \textbf{1-D Pos. Emb.} 
\\
\hline
RM1 & \textbf{0.801}  & 0.796 & \textbf{78.70} & 78.34 & \textbf{0.455} &  0.461 \\
RM2 & \textbf{0.790}  & 0.786 & \textbf{81.22} & 80.94 & \textbf{0.423} & 0.426 \\
RM3 & \textbf{0.775}  & Not Converge & \textbf{83.76} & Not Converge & \textbf{0.382} & Not Converge  \\
 \hline
\end{tabular}}
\label{table:pos_emb}
\end{table}

\begin{figure*}[h!]
  \centering
  \subfloat{
	\begin{minipage}[t]{0.32\textwidth}
	   \centering
	   \includegraphics[width=\textwidth]{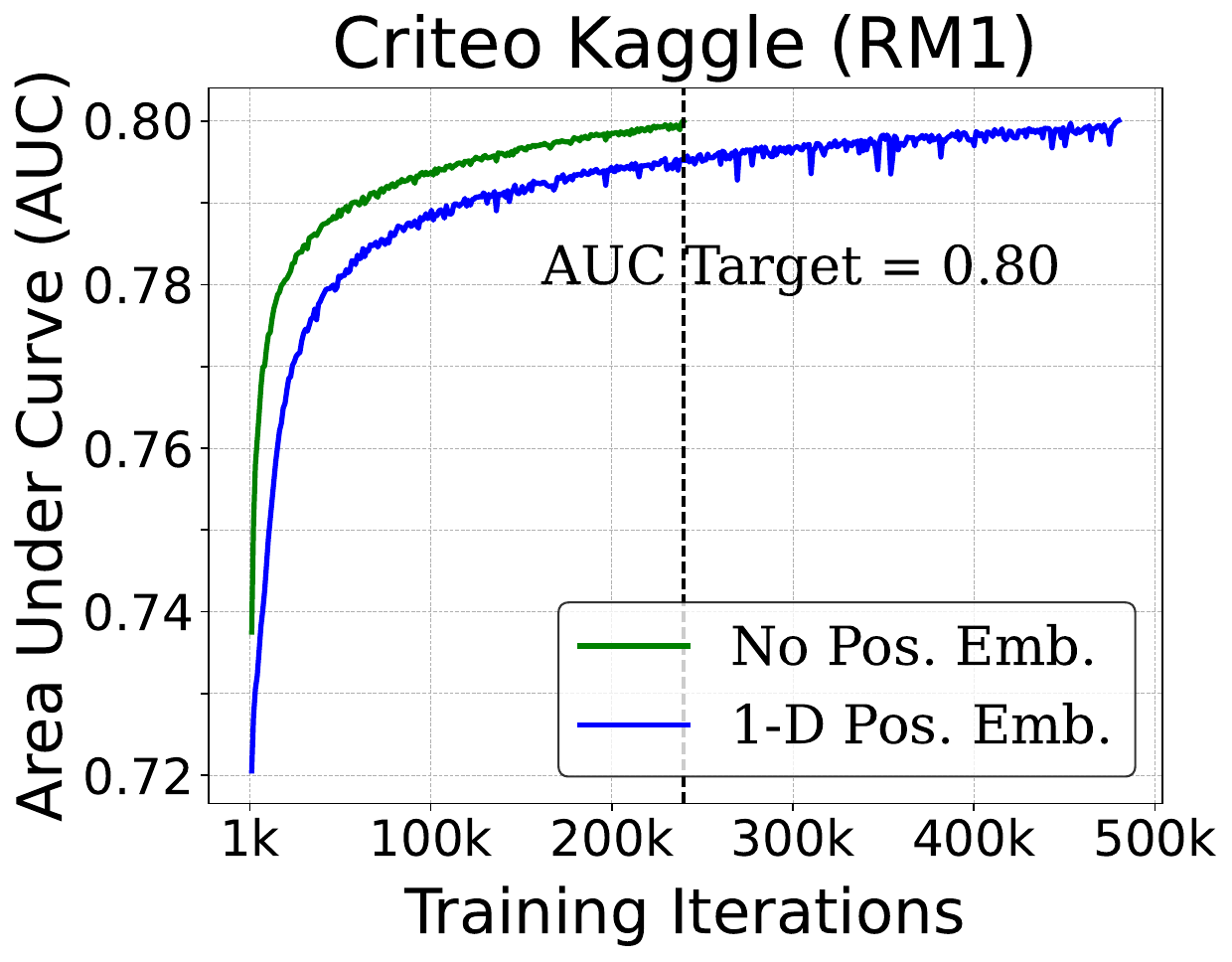}
	\end{minipage}}
  \subfloat{
	\begin{minipage}[t]{0.30\textwidth}
	   \centering
	   \includegraphics[width=\textwidth]{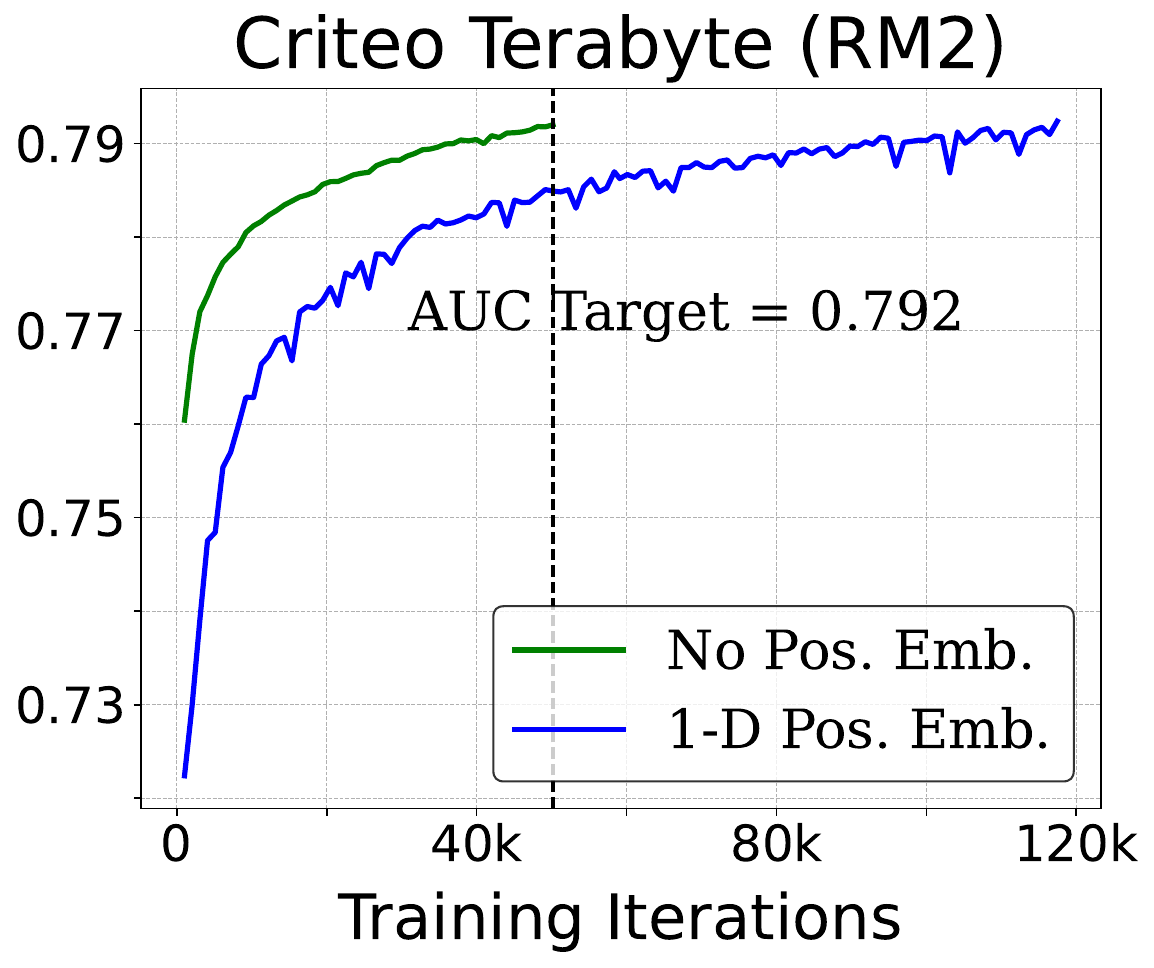}
	\end{minipage}}
  \subfloat{
	\begin{minipage}[t]{0.29\textwidth}
	   \centering
	   \includegraphics[width=1\textwidth]{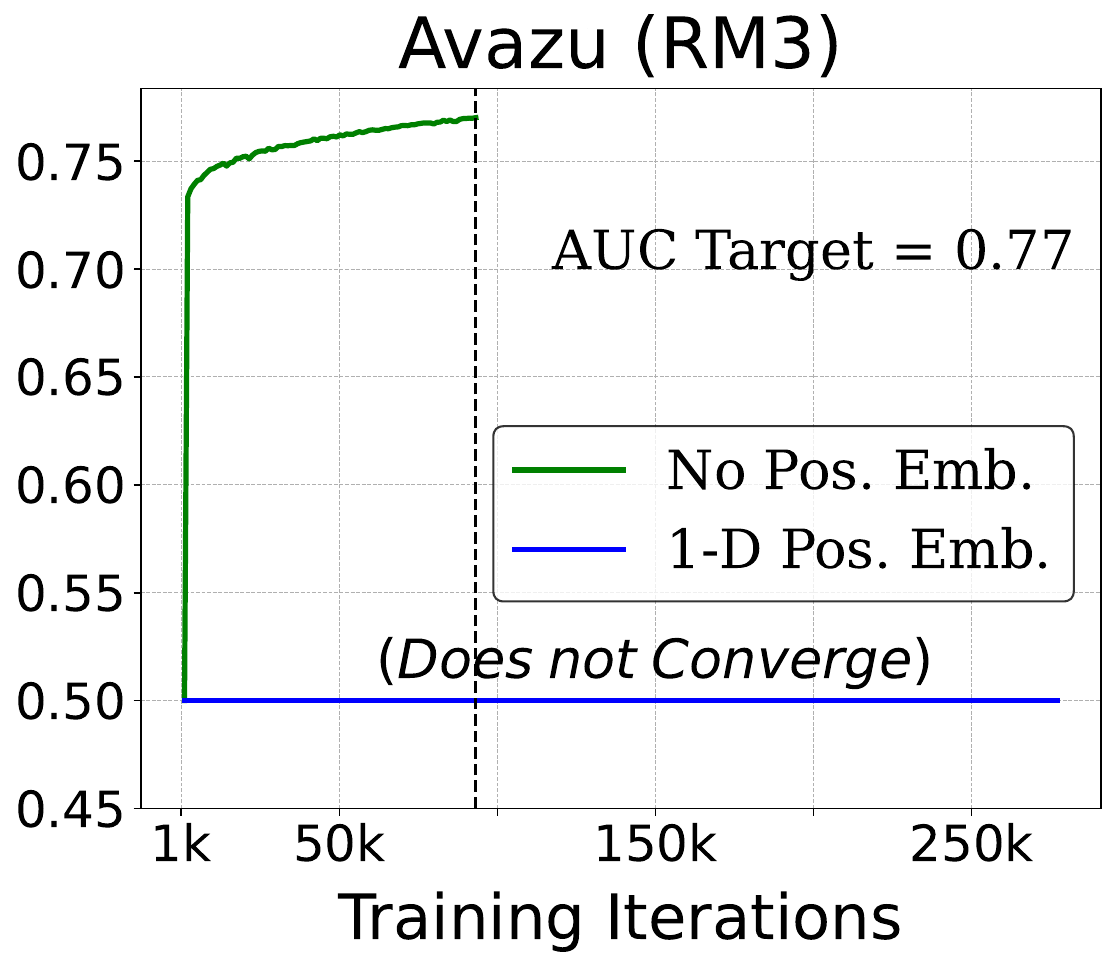}
	\end{minipage}}
\caption{Model convergence with and without positional embedding.}
\label{fig:pos_emb}
\end{figure*}

Figure~\ref{fig:pos_emb} illustrates that recommendation models achieve convergence in significantly fewer training iterations when no spatial information is added to the input feature embeddings. Even the model with 1-dimensional positional embedding requires more training iterations than the baseline DLRM with second-order cross-features to reach the target AUC.

\subsection{\trec{}: Masked Transformer Hyper-parameters Analysis}
\label{subsec:rectra_hyperparameters}

To evaluate the robustness of our \trec{} model, we conducted an extensive study by varying the hyperparameters of the transformer encoder. Table~\ref{table:heper-para} provides an overview of the different hyperparameters considered in this sensitivity analysis.

\begin{table*}[h!]
\centering
\caption{Scaling hyper-parameters of transformer encoder.}
\resizebox{0.8\textwidth}{!}{
\begin{tabular}{c c c  c c  c }
\hline
\textbf{Num Layers} & \textbf{Num Heads} & \textbf{Dropout Ratio} & \textbf{Non-Linear Activation} & \textbf{Hidden Size} & \textbf{Linear Config.}
\\
 \textbf{$N$}& \textbf{$H$} & \textbf{$D$} & \textbf{$A$} & \textbf{$d_{model}$} & \textbf{$d_{ff}$} \\
\hline
\hline
1 & 1 & 0.01 & ReLU & 16 & 128 \\ 
2 & 2 & 0.05 & GeLU & 64 & 512 \\ 
4 & 4 & 0.1 &  &  &  \\
 & 8 & 0.2 &  &  &  \\
 & 16 & 0.3 &  &  &  \\
\hline
\end{tabular}}
\label{table:heper-para}
\vskip -0.2in
\end{table*}

\subsubsection{Number of Layers}
\label{subsubsec:encoder_layers}
\begin{figure*}[h!]
  \centering
  \subfloat{
	\begin{minipage}[t]{0.4\textwidth}
	   \centering
	   \includegraphics[width=\textwidth]{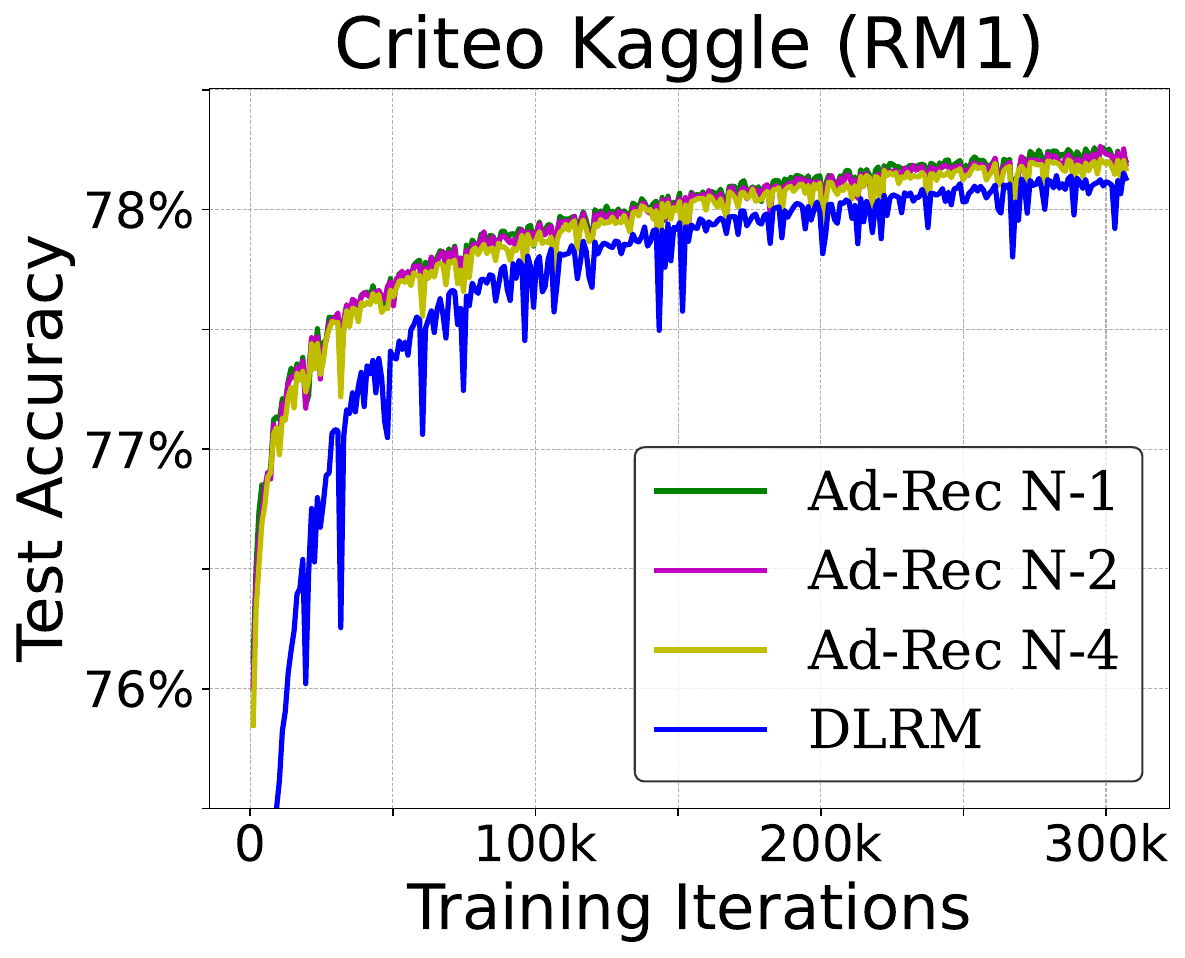}
	\end{minipage}}
  \subfloat{
	\begin{minipage}[t]{0.4\textwidth}
	   \centering
	   \includegraphics[width=\textwidth]{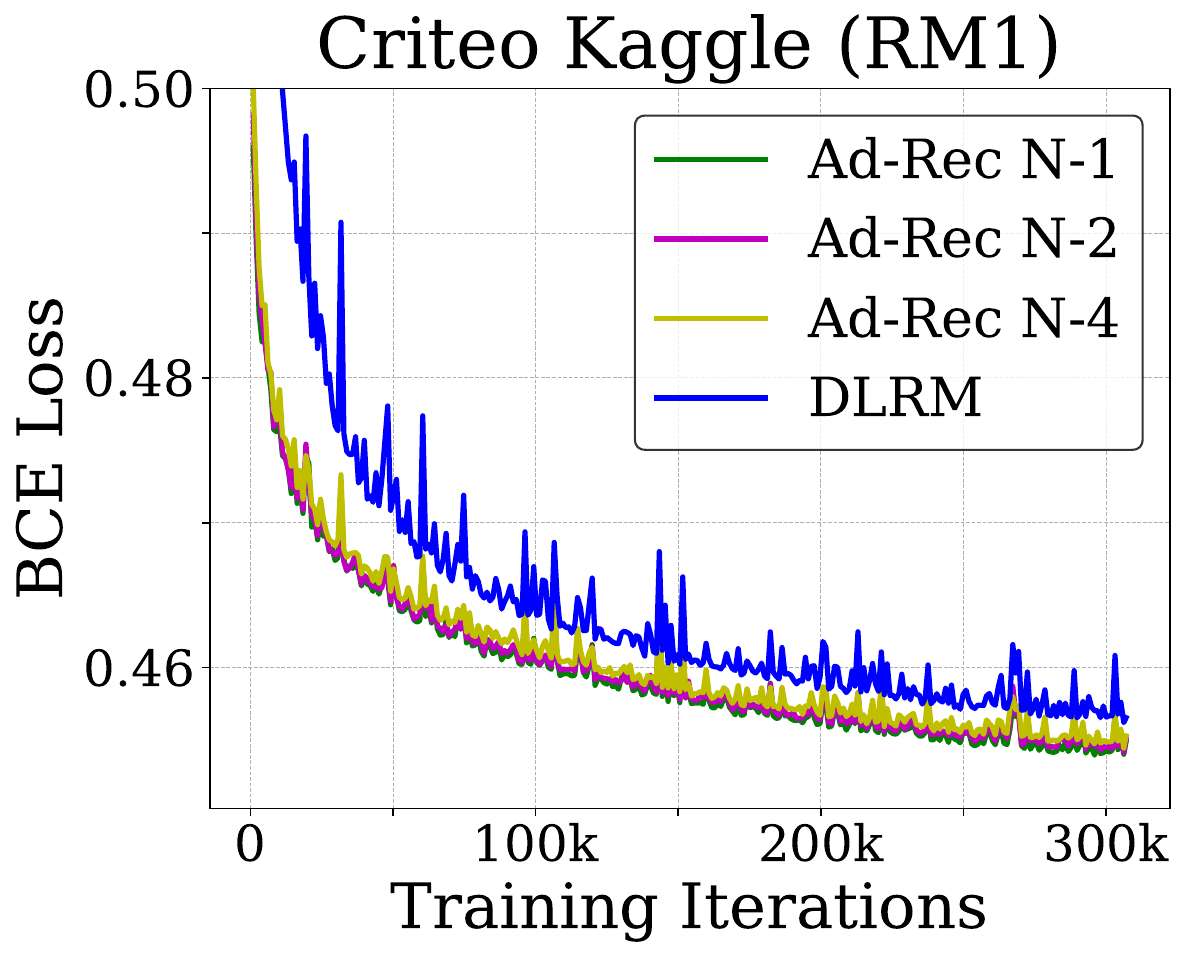}
	\end{minipage}}
\caption{Test Accuracy and BCE Loss with varying layers of \trec{} masked transformer.}
\label{fig:n_layers}
\end{figure*}

We investigated the impact of the number of masked transformer layers $(N)$ in the RM1 model while keeping other hyperparameters constant. Figure~\ref{fig:n_layers} showcases the test accuracy and BCE loss for the RM1 model. Our observations revealed that increasing the number of masked transformer layers did not yield significant benefits due to the smaller sequence length of RM1 (27 features). However, we anticipate that models with a larger number of features and longer sequence lengths would demonstrate improved AUC with more masked transformer layers. This is because higher-order feature interaction becomes more valuable when there are more features. In large-scale industrial datasets, the number of sparse features can reach thousands. For instance, the Meta synthetic dataset\footnote{\hyperlink{https://github.com/facebookresearch/dlrm\_datasets}{https://github.com/facebookresearch/dlrm\_datasets}}, which cannot be used for training due to its synthetic nature, contains 856 sparse features. Nevertheless, it highlights the scale at which \rectra can significantly enhance prediction accuracy.

\subsubsection{Number of Attention Heads}
\begin{figure*}[t]
  \centering
  \subfloat{
	\begin{minipage}[t]{0.24\textwidth}
	   \centering
\includegraphics[width=\textwidth]{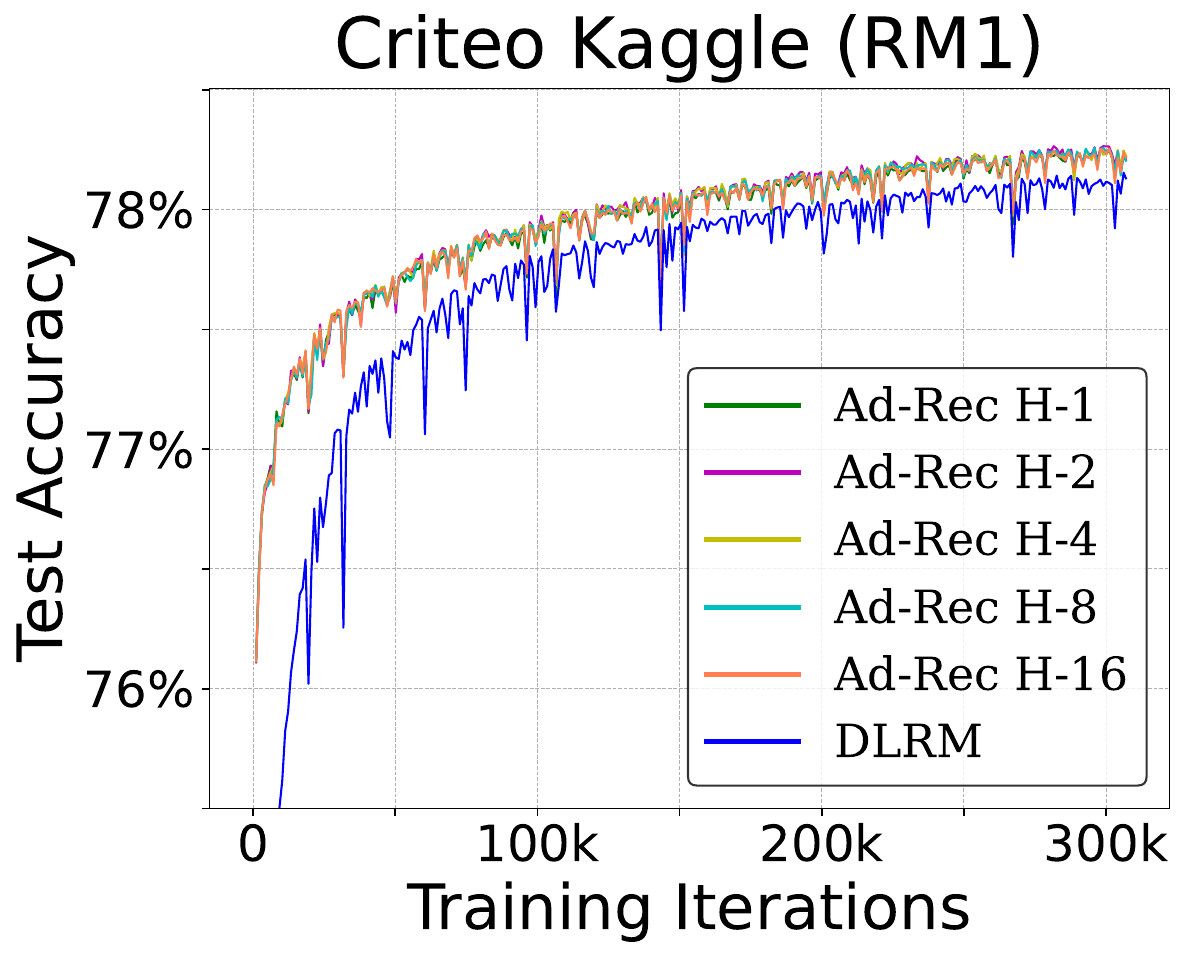}
	\end{minipage}}
  \subfloat{
	\begin{minipage}[t]{0.24\textwidth}
	   \centering
	   \includegraphics[width=\textwidth]{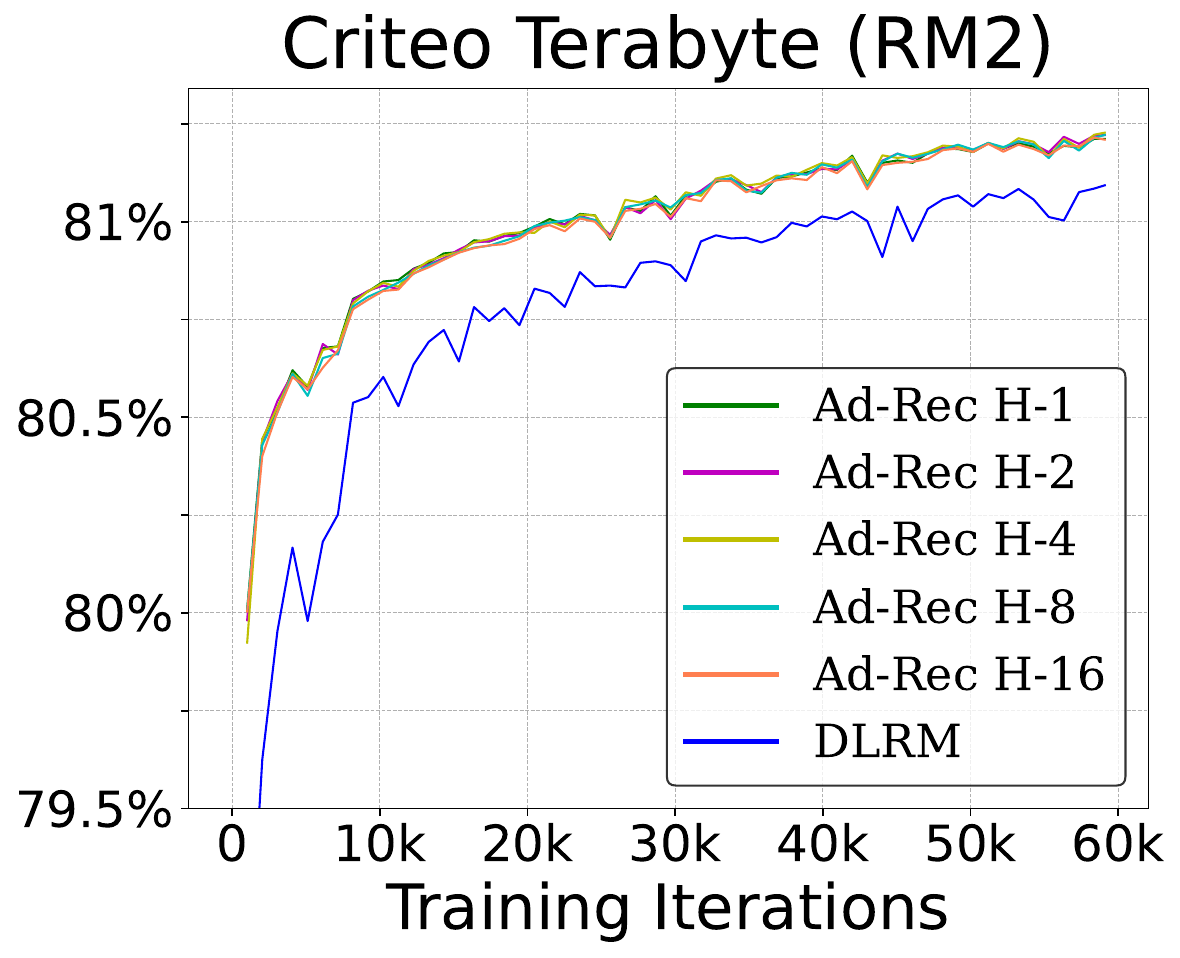}
	\end{minipage}}
 \begin{minipage}[t]{0.24\textwidth}
	   \centering
	   \includegraphics[width=\textwidth]{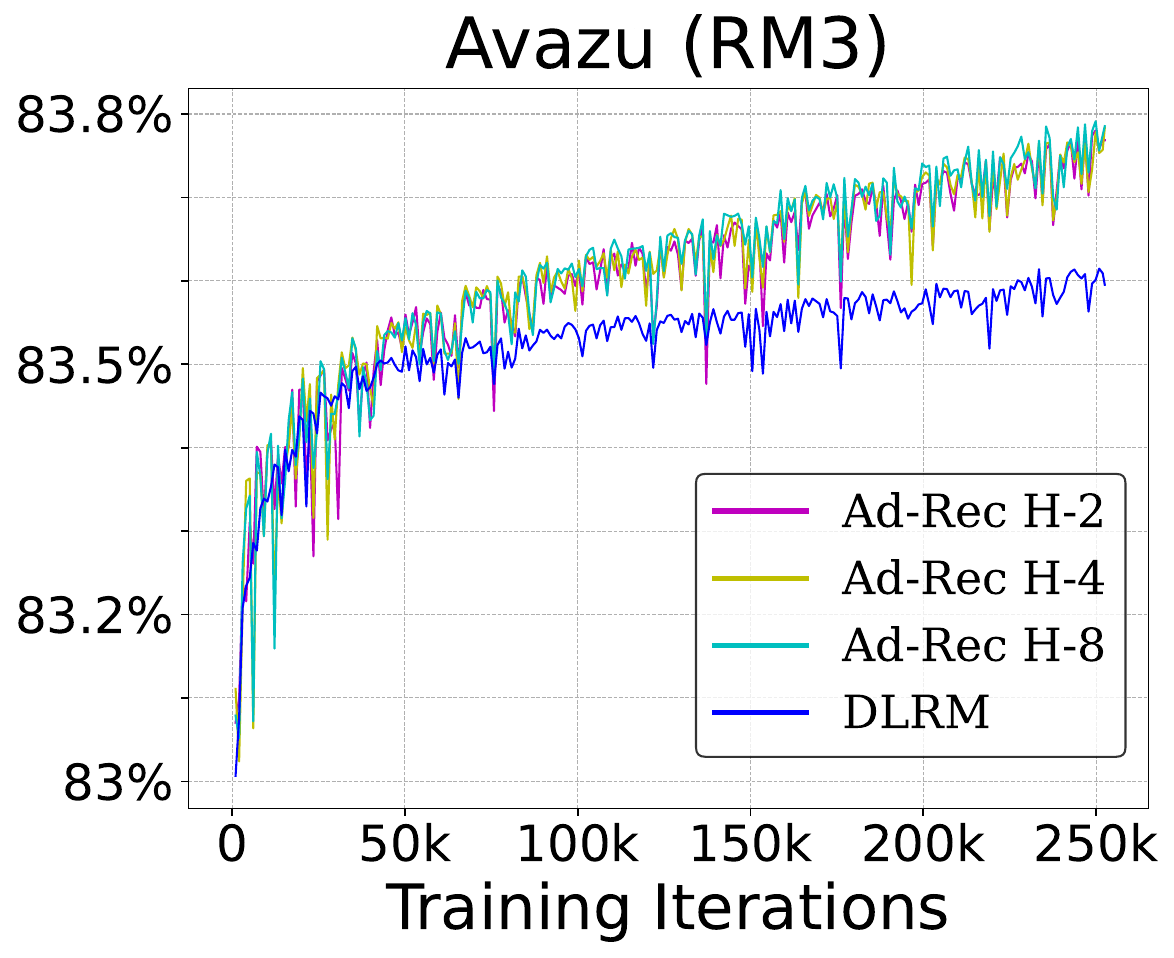}
	\end{minipage}
  \subfloat{
	\begin{minipage}[t]{0.23\textwidth}
	   \centering
	   \includegraphics[width=\textwidth]{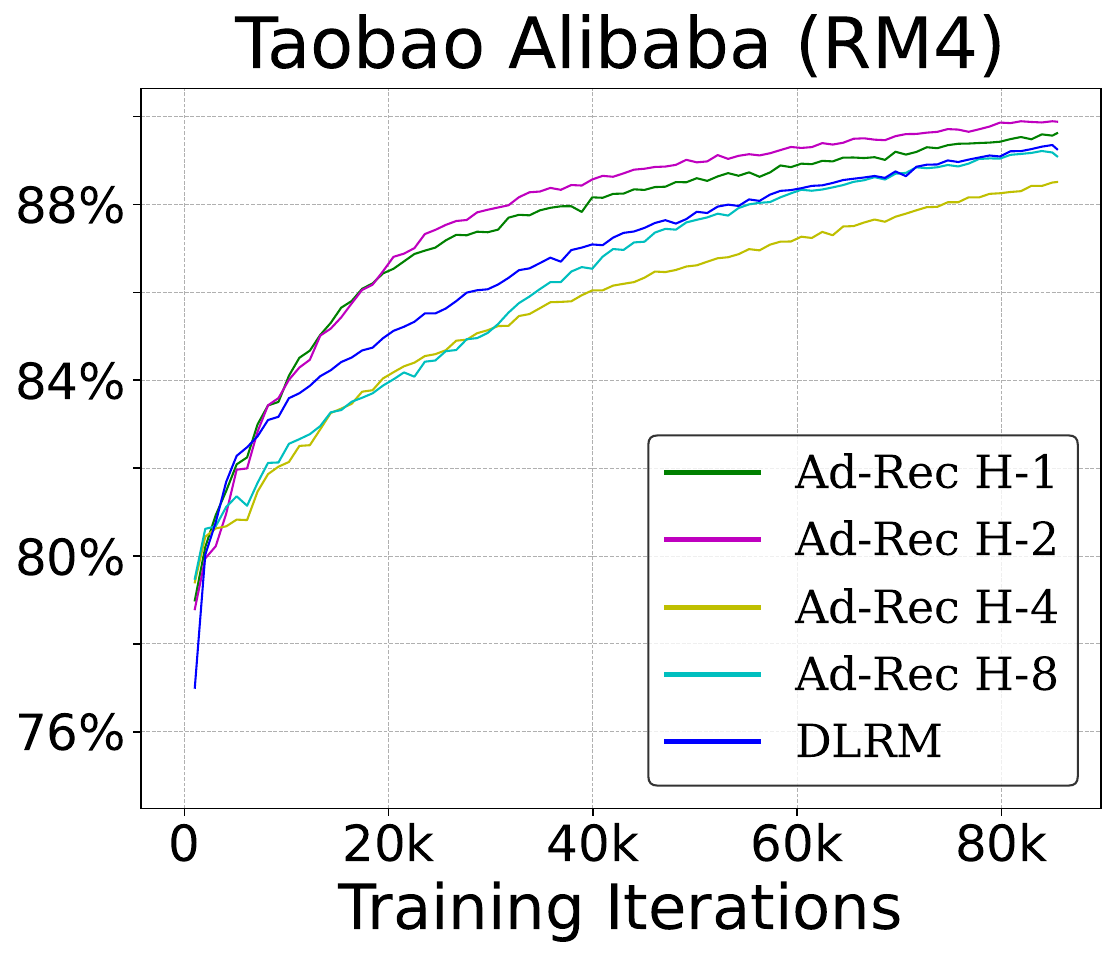}
	\end{minipage}}
\caption{Trends in test accuracy by varying the number of attention heads of masked transformer for RM1, RM2, RM3, and RM4 models. Models with missing head (RM3 and RM4 with 16 heads) means the model could not converge.}
\label{fig:n_heads}
\end{figure*}

We employ masked multi-head attention to enable higher-order feature interactions across multiple subspaces. This technique divides a single embedding vector into multiple heads of the same length, allowing for parallel execution and feature interaction within different subspaces. The number of attention heads $(H)$ varies from 1 to 16 for each model (RM1, RM2, RM3, and RM4). The relationship between the test accuracy and the number of attention heads is illustrated in Figure~\ref{fig:n_heads}. Notably, the computational complexity remains unchanged as all the heads are concatenated at the end, and each head operates on a portion of the embedding vector.

Our observations indicate that models with an intermediate number of heads consistently converge and yield higher accuracy. In contrast, the RM3 model fails to converge when using a single-head model, emphasizing the importance of feature interaction in multiple subspaces. Furthermore, when the number of heads equals the length of the embedding vector $D$ (i.e., $H = D$), the feature interaction becomes excessively fine-grained, leading to model divergence. For the RM3 and RM4 models, it was observed that neither model converged when using $(H=16)$ and $(D=16)$, reinforcing the need for an appropriate balance in the number of attention heads to achieve optimal performance.

\subsubsection{Dropout Ratio}
The masked transformer architecture incorporates a residual connection, depicted in Figure~\ref{fig:transformer_interaction}, to ensure effective information flow and gradient propagation. Additionally, a dropout mechanism prevents overfitting by randomly replacing input features with random features. To investigate the impact of the dropout ratio on model predictions, we varied the ratio across all models, ranging from 0.01 to 0.3.

\begin{figure*}[h!]
  \centering
  \subfloat{
	\begin{minipage}[t]{0.24\textwidth}
	   \centering
\includegraphics[width=\textwidth]{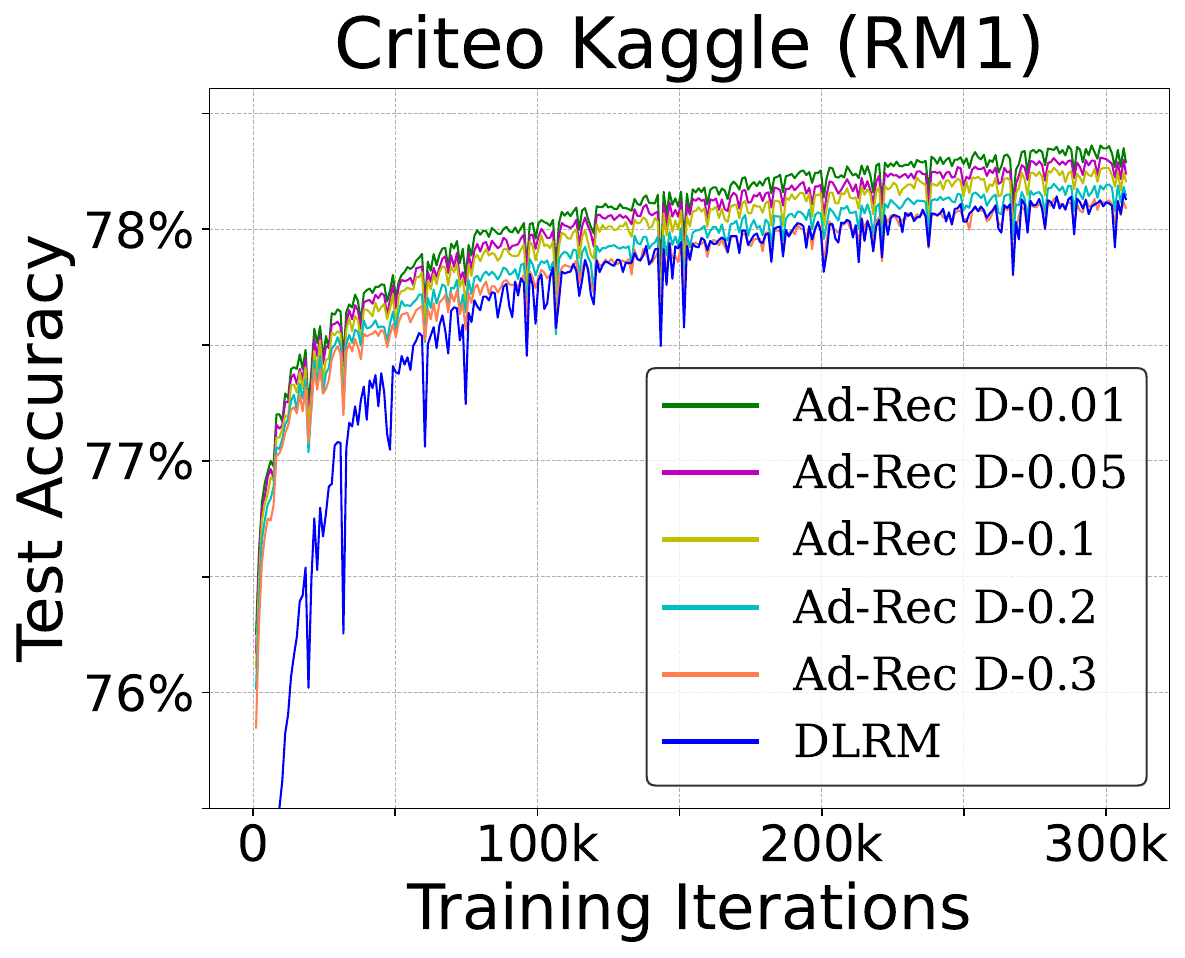}
	\end{minipage}}
  \subfloat{
	\begin{minipage}[t]{0.24\textwidth}
	   \centering
	   \includegraphics[width=\textwidth]{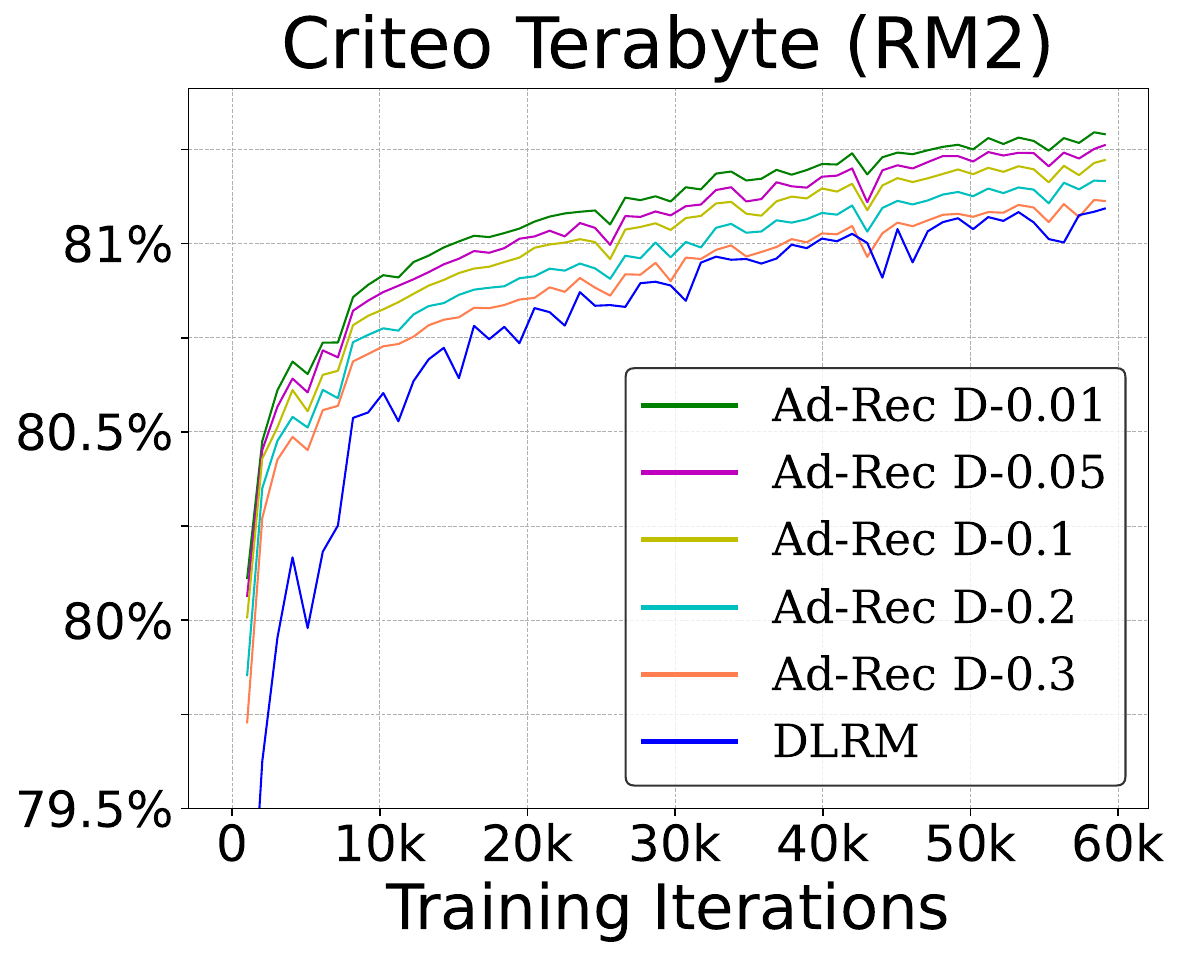}
	\end{minipage}}
 \begin{minipage}[t]{0.24\textwidth}
	   \centering
	   \includegraphics[width=\textwidth]{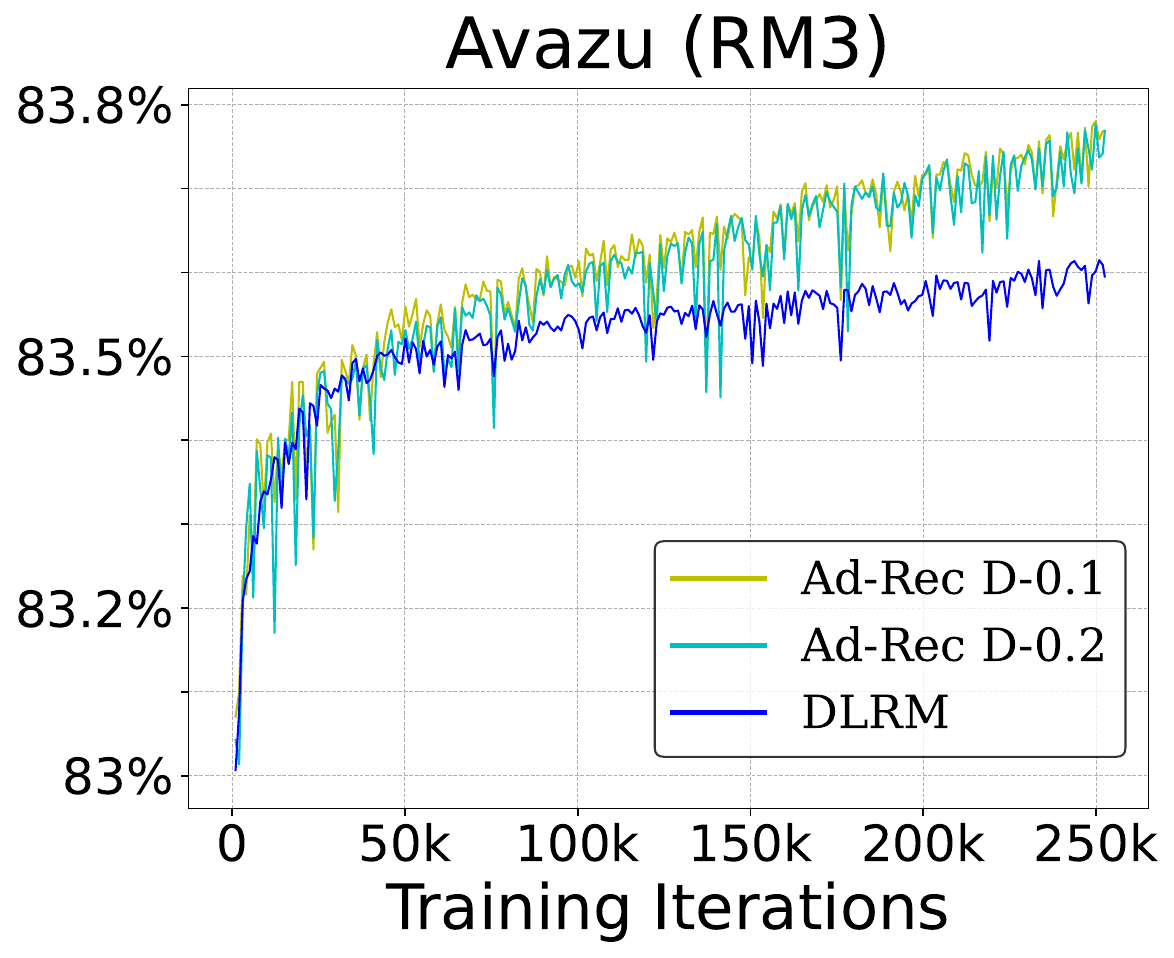}
	\end{minipage}
  \subfloat{
	\begin{minipage}[t]{0.23\textwidth}
	   \centering
	   \includegraphics[width=\textwidth]{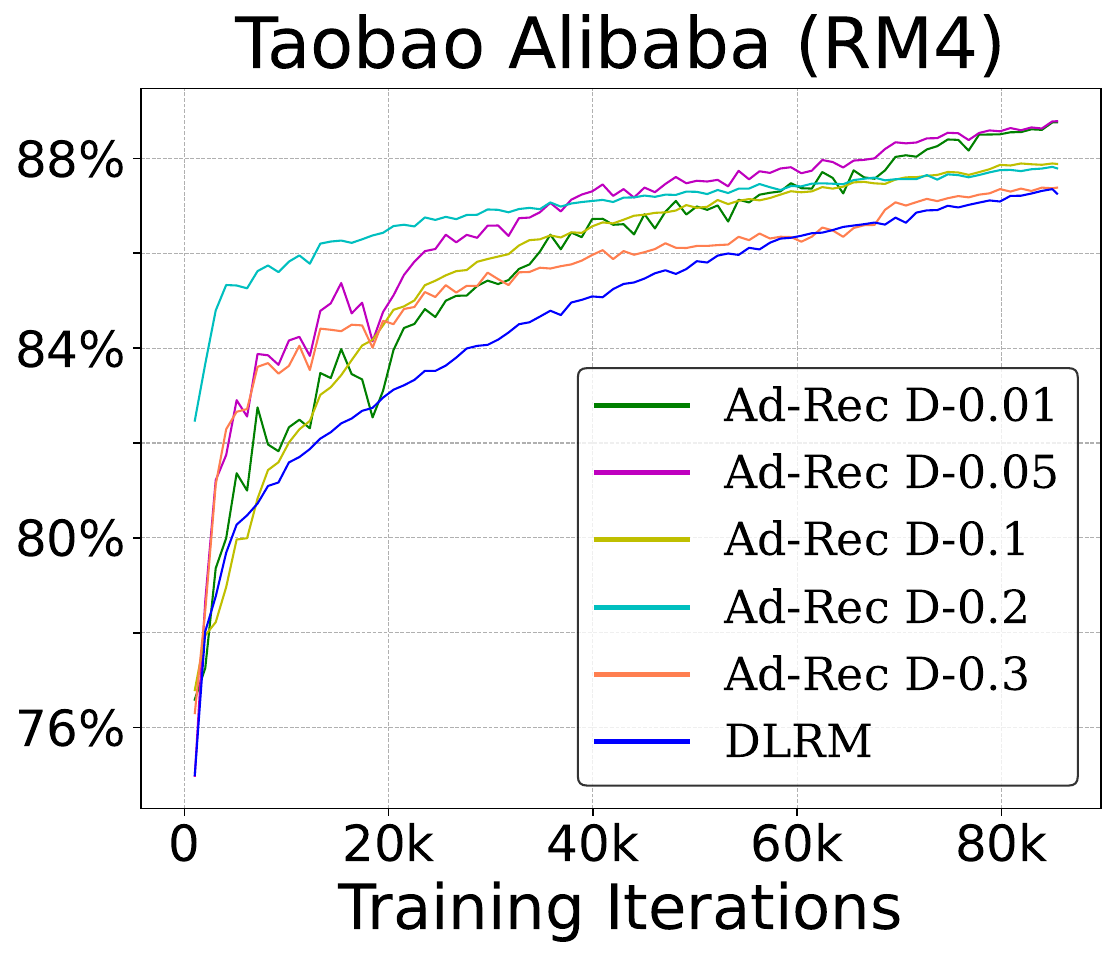}
	\end{minipage}}
\caption{Trends in test accuracy by varying dropout ratio of residual connection in masked transformer for RM1, RM2, RM3, and RM4 model. Models with missing dropout values (RM3 with a dropout of 0.01 and 0.05) mean the model could not converge.}
\label{fig:dropout}
\end{figure*}

Figure~\ref{fig:dropout} showcases the relationship between the dropout ratio and test accuracy. Our findings consistently demonstrate that lower dropout values (0.01 - 0.05) yield superior predictions across all models. These smaller dropout ratios enable cross-features to incorporate with the original raw features, facilitating improved learning of implicit interactions. As the dropout ratio increases, the test accuracy gradually declines, eventually approaching the performance of the baseline DLRM model when the dropout ratio reaches 0.3.

The diminishing test accuracy with higher dropout ratios can be attributed to introducing more random features into the cross-features. These additional random features may disrupt the underlying patterns and relationships in the data, negatively impacting model performance. This finding aligns with previous research~\citep{autoint}, which utilizes raw features without dropout in multi-head attention-based cross-features to learn higher-order interactions. However, in the case of \trec{}, a smaller dropout ratio is employed during training to enhance generalization, while dropout is removed during inference for optimal performance.

\subsubsection{Non-Linear Activation}
\begin{figure*}[h!]
  \centering
  \subfloat{
	\begin{minipage}[t]{0.24\textwidth}
	   \centering
\includegraphics[width=\textwidth]{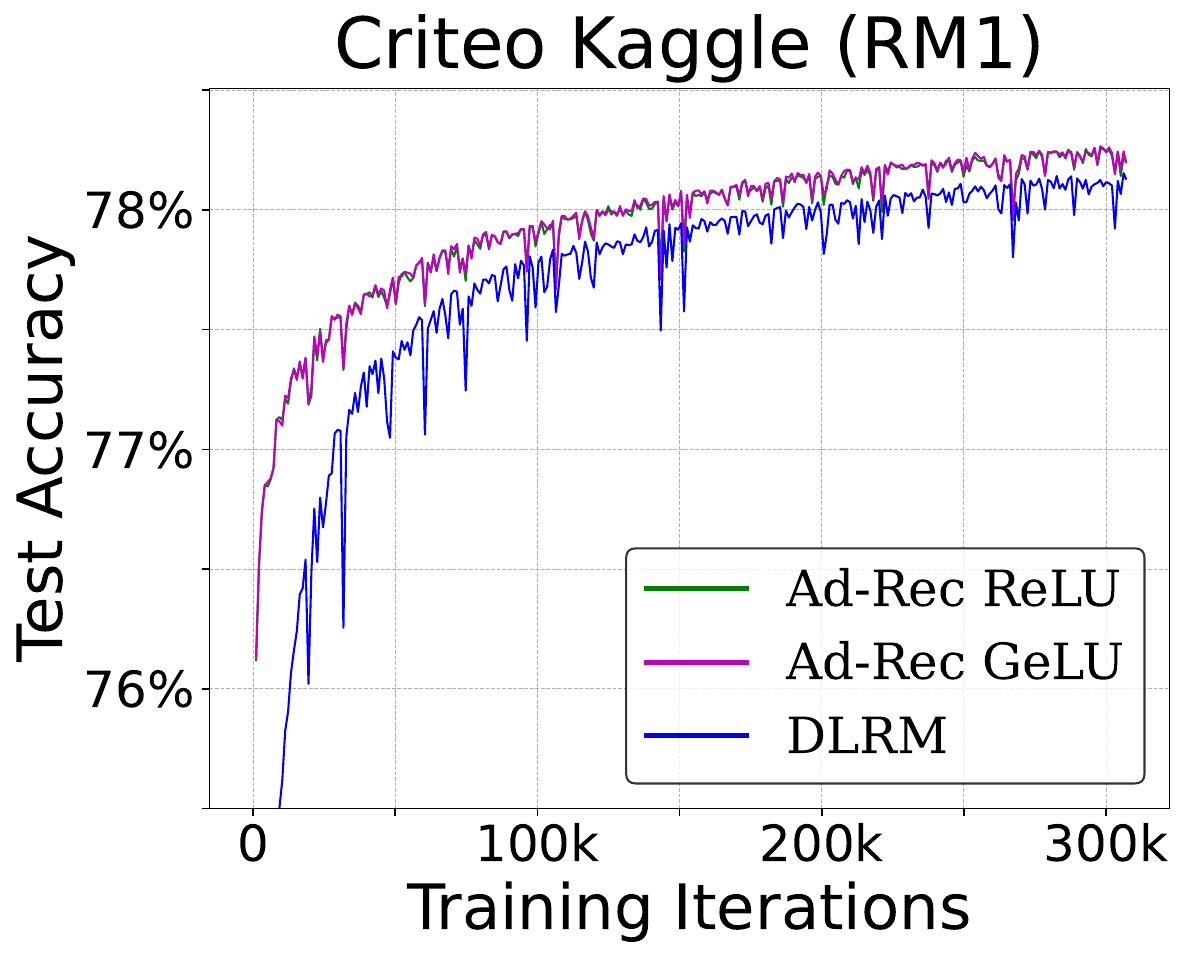}
	\end{minipage}}
  \subfloat{
	\begin{minipage}[t]{0.24\textwidth}
	   \centering
	   \includegraphics[width=\textwidth]{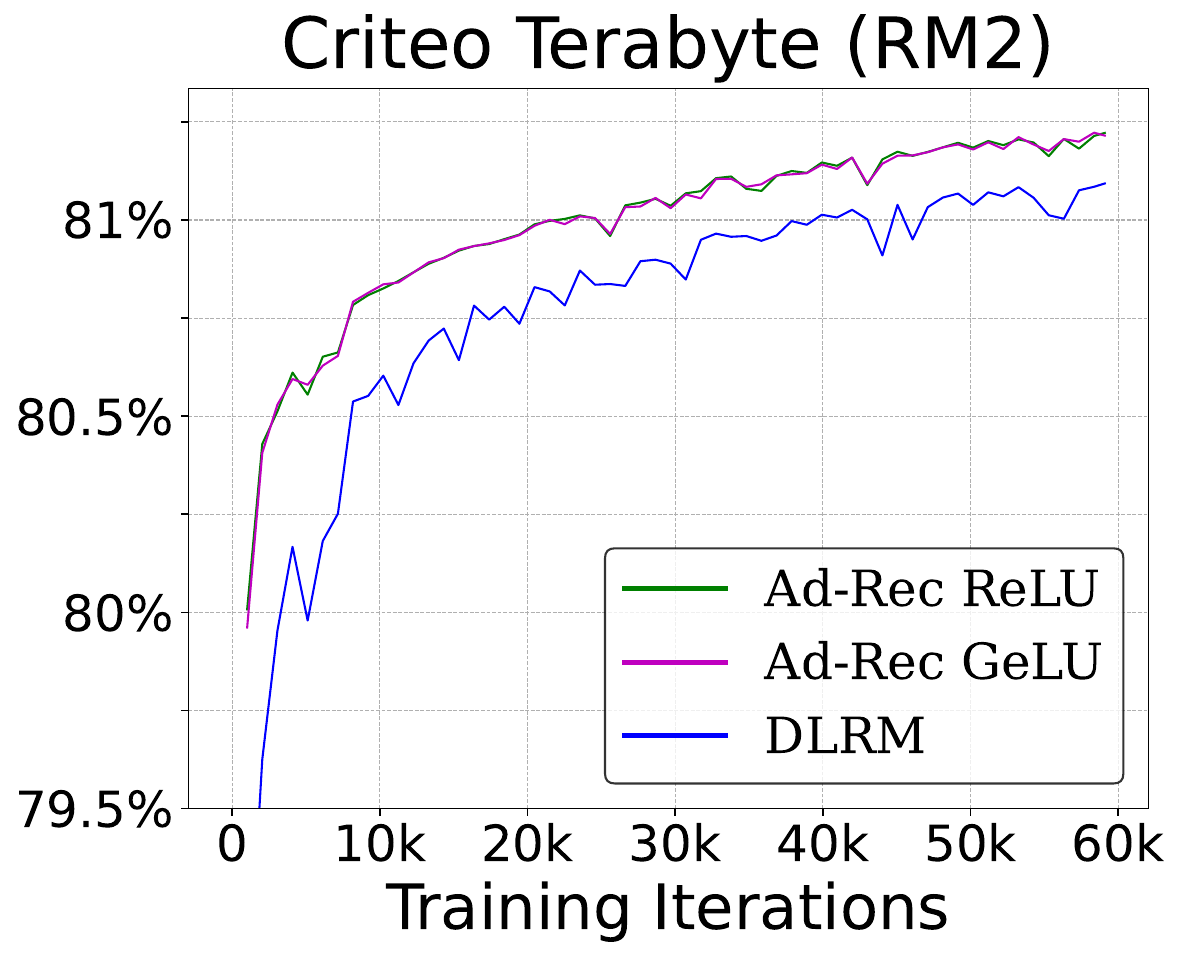}
	\end{minipage}}
 \begin{minipage}[t]{0.24\textwidth}
	   \centering
	   \includegraphics[width=\textwidth]{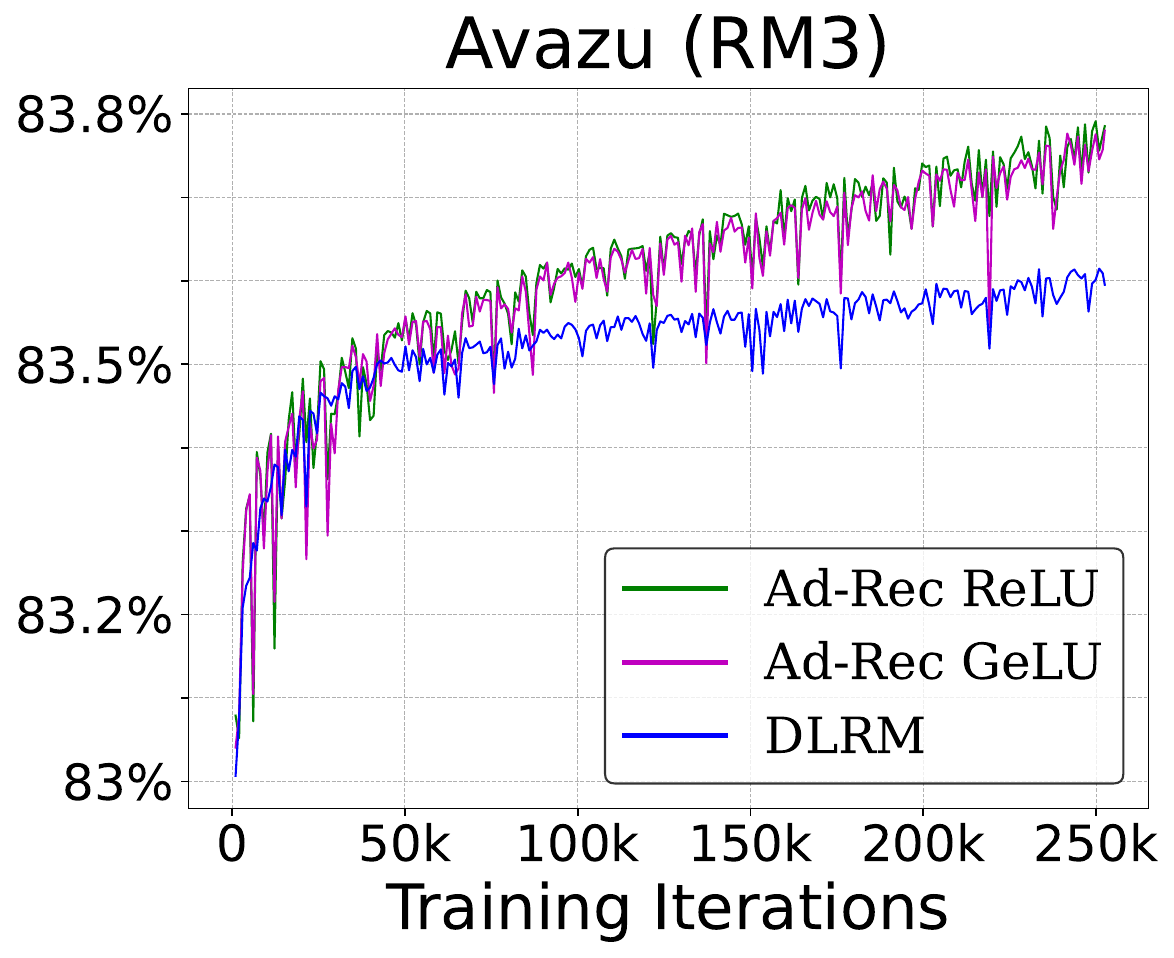}
	\end{minipage}
  \subfloat{
	\begin{minipage}[t]{0.23\textwidth}
	   \centering
	   \includegraphics[width=\textwidth]{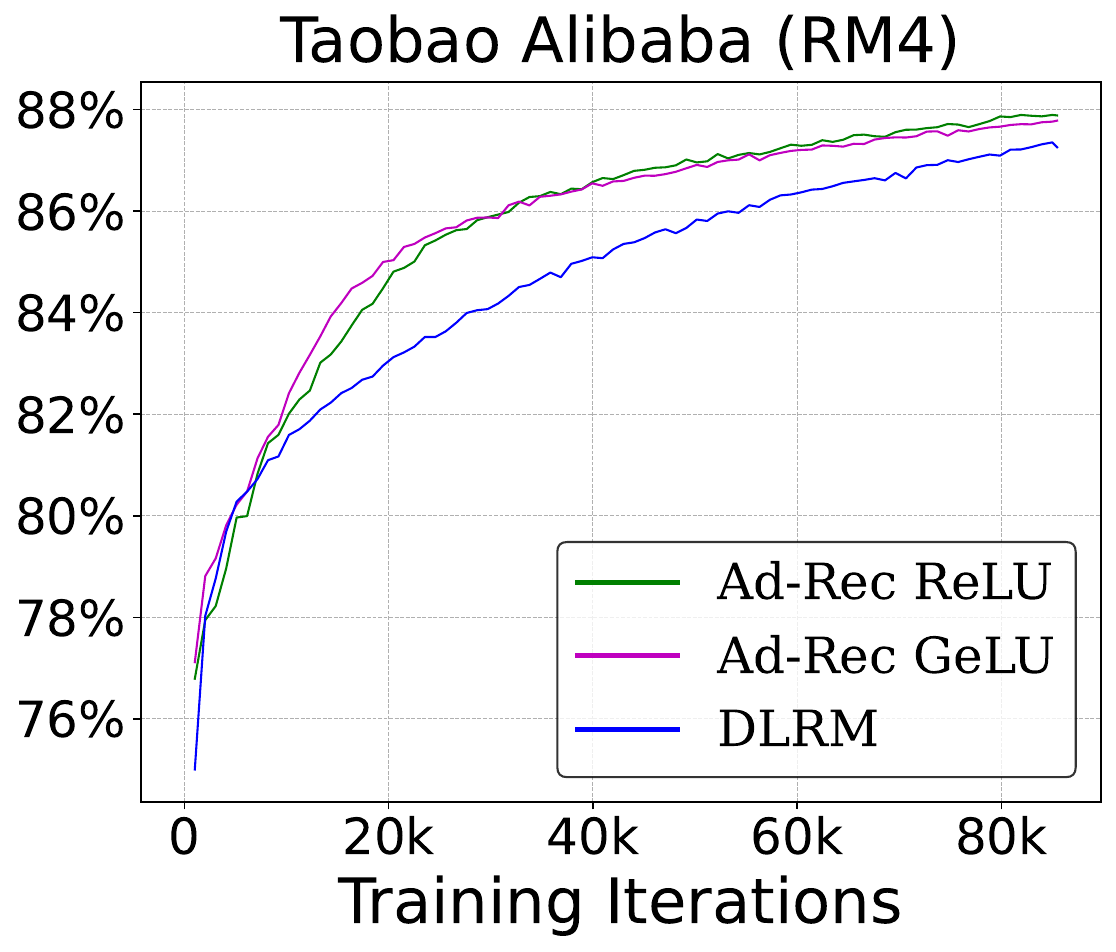}
	\end{minipage}}
\caption{Trends in test accuracy by changing the non-linear activation of a feed-forward network of the masked transformer. It does not have any effect on the model quality.}
\label{fig:activation}
\end{figure*}

Figure~\ref{fig:activation} presents the investigation into the impact of non-linear activation functions on test accuracy. Surprisingly, transitioning from the default ReLU activation to GeLU activation does not yield any noticeable effect on the test accuracy of the models. Regardless of the chosen non-linear activation function, all models across different datasets converge to the same point.

This finding suggests that the specific type of non-linearity employed in the activation function does not significantly influence the extraction of cross-features. It indicates that the masked transformer architecture is robust to different non-linearities, and the models can effectively capture and learn the underlying interactions between features regardless of the specific activation function used.

\subsection{Attention Heads Masking Analysis}
\label{subsec:masked_attention_heads}
\begin{figure*}[t]
  \centering
  \includegraphics[width=0.7\textwidth]{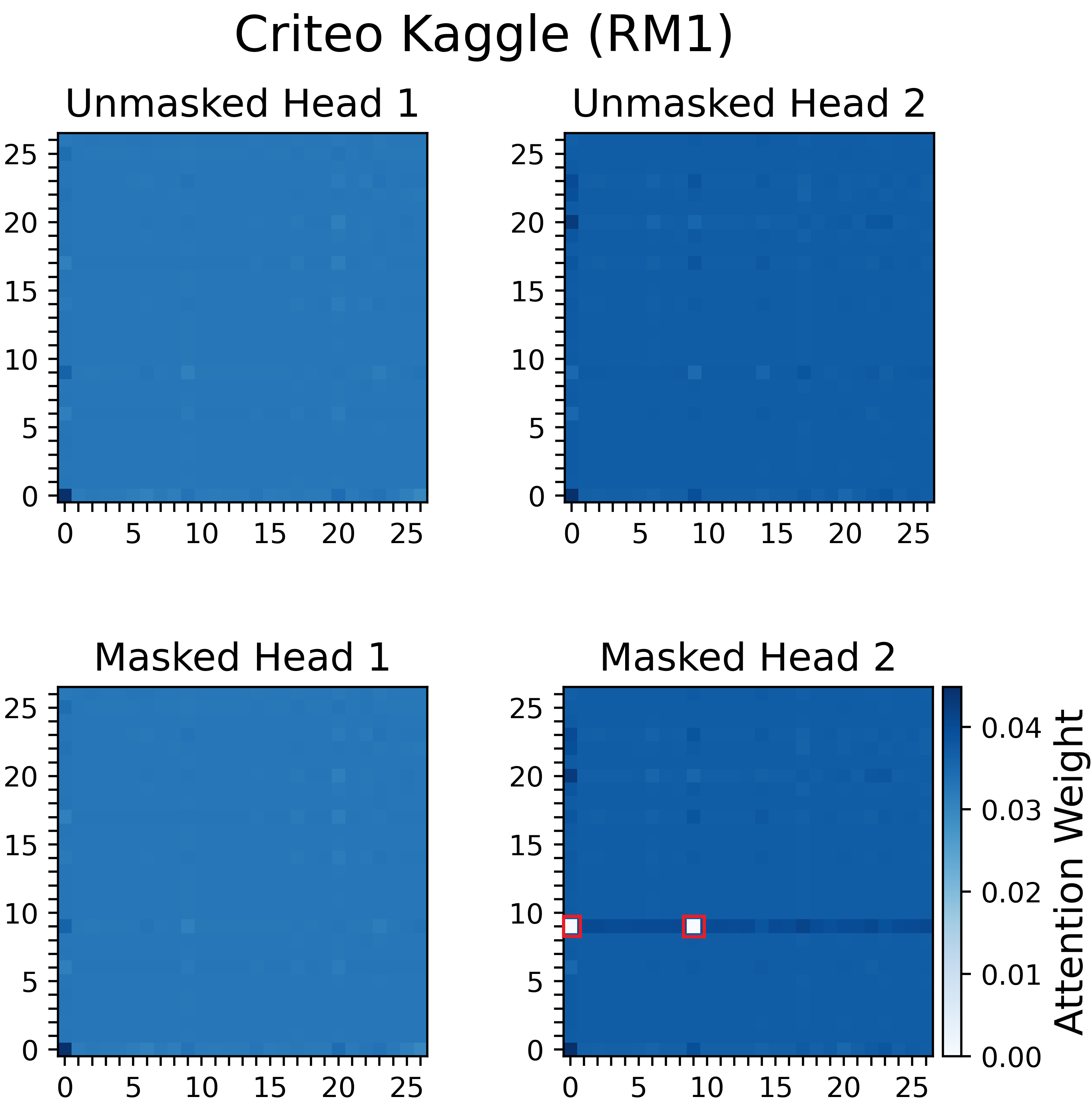}
\caption{Attention Weights for \textbf{RM1} model across 2 heads. The unmasked head contains original attention weights, while the masked head contains attention weights after masking. X and Y axes contain 27 features with 0 as dense feature vectors while others are sparse feature vectors. Highlighted features are the masked features that are irrelevant.}
\label{fig:kaggle_attention_weights}
\vspace{-0.1in}
\end{figure*}

\begin{figure*}[t]
  \centering
  \includegraphics[width=1\textwidth]{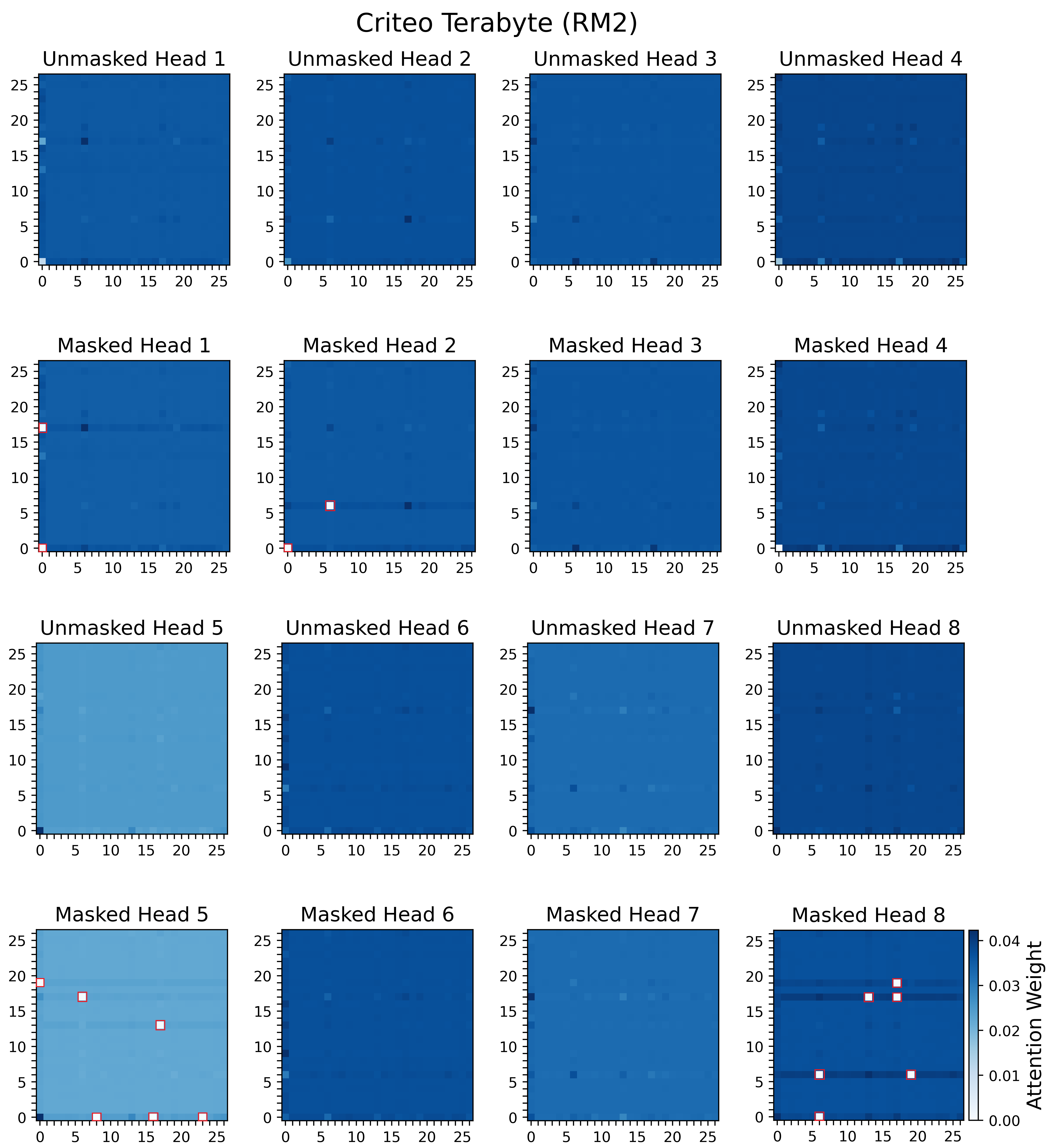}
\caption{Attention Weights for \textbf{RM2} model across 8 heads. The unmasked head contains original attention weights, while the masked head contains attention weights after masking. X and Y axes contain 27 features with 0 as dense feature vectors, while others are sparse feature vectors. Highlighted features are the masked features that are irrelevant.}
\label{fig:terabyte_attention_weights}
\end{figure*}

\begin{figure*}[t]
  \centering
  \includegraphics[width=0.7\textwidth]{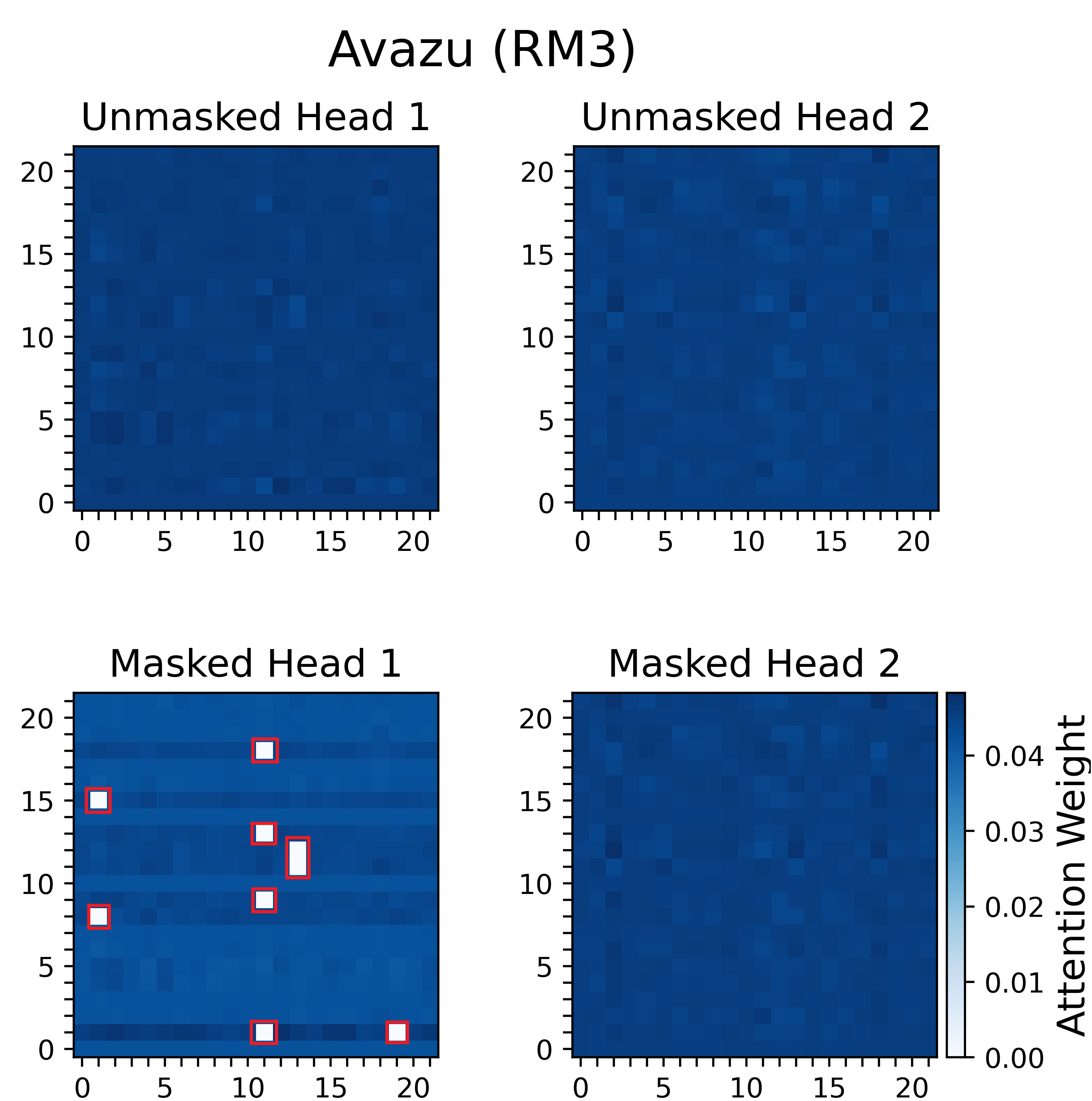}
\caption{Attention Weights for \textbf{RM3} model across 2 heads. The unmasked head contains original attention weights, while the masked head contains attention weights after masking. X and Y axes contain 22 features with 0 as dense feature vectors, while others are sparse feature vectors. Highlighted features are the masked features that are irrelevant.}
\label{fig:avazu_attention_weights}
\vspace{-2ex}
\end{figure*}

We randomly sampled test samples from each dataset to examine the impact of masking on each attention head and its role in enhancing predictions by eliminating irrelevant features. We then plotted the attention weights for each attention head with and without a mask. Figures~\ref{fig:kaggle_attention_weights},\ref{fig:terabyte_attention_weights}, and\ref{fig:avazu_attention_weights} showcase the attention weights for the RM1, RM2, and RM3 models, respectively. These plots provide insights into how different attention heads learn feature interactions in multiple subspaces and how attention weights vary across different attention heads. By employing different mask values for each attention head, masking selectively masks out attention weights of irrelevant features, redistributing the attention weight across other relevant features accordingly.

\begin{figure}
\begin{center}
\centerline{\includegraphics[width=0.6\columnwidth]{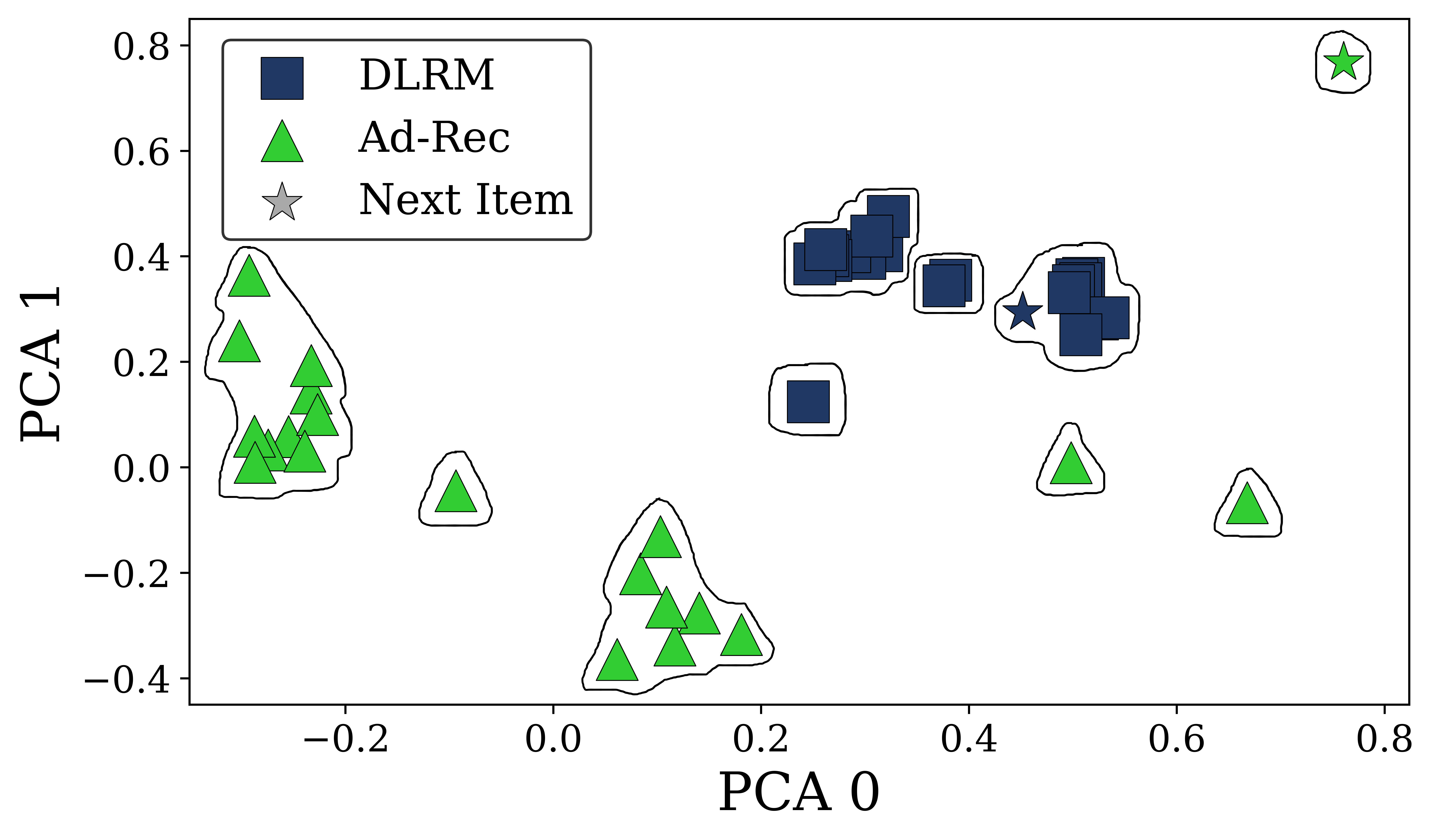}}
\caption{Principal Component Analysis (PCA) to represent high-dimensional sequential user activity in 2-dimensional space. This compares DLRM and \trec generated embedding vectors for the RM4 model. Note that the next item (denoted by $\star$) is a \emph{negative sample}. Thus, a \emph{large Euclidean distances} between the next item and sequential interaction clusters indicate a higher quality model.}
\label{fig:pca}
\end{center}
\vskip -0.2in
\end{figure}

\paragraph{Sequential Embedding Vectors' Feature Interaction:}

The quality of sequence embedding vectors $(Z)$ plays a crucial role in the prediction quality of sequential recommendation. To better understand the patterns within these vectors, we utilize Principal Component Analysis (PCA) to map the embedding vectors (with a width of 16) into a 2-dimensional space. In this space, similar interactions are grouped closely together, while dissimilar interactions are distanced apart. The proximity of the next item's placement to historical item interactions indicates a higher probability of the user clicking on that item. Figure~\ref{fig:pca} depicts the PCA plot for sequential embedding vectors generated by both feature interaction methods, as explained in Section~\ref{sec:trec_analysis}. Each method generates 21 embedding vectors, representing the user's historical interactions, with the next item represented by a star symbol $(\star)$.

In the plot, closely related interactions form clusters, and the relative Euclidean distance between clusters signifies their correlation. For DLRM-based feature interaction, we observe that all sequential interactions are clustered together, resulting in four closely located clusters in the Euclidean space. On the other hand, \trec-based feature interaction generates more distinct clusters based on the type of user interactions. The sequential layer predicts the likelihood of the user clicking on the next item vector based on the placement of the next item embedding in relation to the user's sequential embedding vectors.

Interestingly, we notice that \trec-based feature interaction generates an embedding vector for the next item that is situated far away from previous user interactions, indicating that the next item does not have a strong connection to the user's sequential interactions. Conversely, for DLRM-based feature interaction, the next item is located within the cluster, suggesting a close relationship with most previous interactions. The placement of \trec-generated sequence vectors and subsequent item vectors aligns with the ground truth, as the input represents a negative sample.

\subsection{Scaling laws of Recommendation Models}
\label{subsec:scaling_laws}

To investigate the scaling laws of recommendation models, we conducted experiments by scaling various model components. Table~\ref{table:scaling_models} provides an overview of the scaled components and their corresponding configurations, along with the model size in terms of parameters. This analysis allows us to assess the impact of scaling on model quality.

Previous research~\citep{scaling_laws} has explored the scaling laws for non-sequential recommendation models, particularly focusing on Click-Through Rate (CTR) in DLRM-style models. Their findings revealed that increasing the model size did not significantly enhance accuracy, while training on more data led to slight improvements.

In contrast, Figure~\ref{fig:scaling_laws} demonstrates the behaviour of model loss as different components, including the \trec loss, are scaled. Interestingly, even with fewer parameters (approximately half), \trec outperforms other models regarding convergence speed. This emphasizes the significance of higher-order feature interaction, eliminating irrelevant features, and addressing covariate shifts in improving the representation of input features. Merely scaling existing components or increasing the size of training datasets does not yield comparable results. The success of \trec opens up new avenues for research in recommendation model architecture.

\begin{table*} [h!]
\centering
\caption{Scaling Recommendation Models Components}
\scriptsize
\resizebox{1\textwidth}{!}{
\begin{tabular}{c c c  c c  c }
\hline
\multirow{2}{*}{\textbf{Model}} & \textbf{Scaling} & \textbf{Sparse} &
\multicolumn{2}{c}{\textbf{Neural Network Configuration}} & \textbf{Model}
\\
 & \textbf{Component} & \textbf{Dimension} & \textbf{Bottom MLP} & \textbf{Top MLP} & \textbf{Parameters} \\
\hline
\hline
RM1 & (N/A)~DLRM & 16 & 13-512-256-64-16 & 512-256-1 & 540.7M  \\
RM1 & (N/A)~\rectra & 16 & 13-512-256-64-16 & 512-256-1 & 540.8M \\
\hline
RM1 & Sparse Emb. Dim & 32 & 13-512-256-64-32 & 512-256-1 & 1.08B \\
RM1 & Top MLP & 16 & 13-512-256-64-16 & 1024-768-512-256-1 & 542.1M \\
RM1 & Bottom MLP & 16 & 13-1024-768-512-256-128-64-16 & 512-256-1 & 542M \\
RM1 & All Comp. & 32 & 13-1024-768-512-256-128-64-32 & 1024-768-512-256-1 & 1.08B \\
\hline
\hline
RM2 & (N/A)~DLRM & 64 & 13-512-256-64 & 512-512-256-1 & 2.7B \\
RM2 & (N/A)~\rectra & 64 & 13-512-256-64 & 512-512-256-1 & 2.701B\\
\hline
RM2 & Sparse Emb. Dim & 128 & 13-512-256-128 & 512-512-256-1 & 5.399B \\
RM2 & Top MLP & 64 & 13-512-256-64 & 1024-768-512-512-256-128-1 & 2.701B\\
RM2 & Bottom MLP & 64 & 13-1024-768-512-256-128-64 & 512-512-256-1 & 2.701B\\
RM2 & All Comp. & 128 & 13-1024-768-512-256-128 & 1024-768-512-512-256-128-1 & 5.401B \\
\hline 
\hline
RM3 & (N/A)~DLRM & 16 & 1-512-256-64-16 & 512-256-1 & 150.24M \\
RM3 & (N/A)~\rectra & 16 & 1-512-256-64-16 & 512-256-1 &  150.37M\\
\hline
RM3 & Sparse Emb. Dim & 32 & 1-512-256-64-32 & 512-256-1 & 300.08M \\
RM3 & Top MLP & 16 & 1-512-256-64-16 & 1024-768-512-256-128-64-1 & 151.59M \\
RM3 & Bottom MLP & 16 & 1-1024-768-512-256-128-64-16 & 512-256-1 & 151.44M \\
RM3 & All Comp. & 32 & 1-1024-768-512-256-128-64-32 & 1024-768-512-256-128-64-1 & 302.64M \\
\hline
\hline
RM4 & (N/A)~DLRM & 16 & 1-16 & 15-15 & 82.55M \\
RM4 & (N/A)~\rectra & 16 & 1-16 & 15-15 & 82.57M \\
\hline
RM4 & Sparse Emb. Dim & 32 & 1-32 & 15-15 & 165.10M \\
RM4 & Top MLP & 16 & 1-16 & 32-15-15 & 82.55M \\
RM4 & Bottom MLP & 16 & 1-8-16 & 15-15 & 82.55M \\
RM4 & All Comp. & 32 & 1-16-32 & 32-15-15 & 165.10M \\
\hline
\hline
\end{tabular}}
\label{table:scaling_models}
\vspace{-3ex}
\end{table*}

\end{document}